\documentclass[11pt]{article}

\usepackage[T1]{fontenc}
\usepackage[utf8]{inputenc}
\usepackage[english]{babel}
\usepackage{amsfonts, amssymb, amsmath}
\usepackage{bm}
\usepackage{xspace}
\usepackage{xparse}
\usepackage{commath}
\usepackage{caption}
\usepackage{subcaption}

\usepackage{tikz}
\usepackage{todonotes}
\usepackage{graphicx}

\usepackage{placeins}
\usepackage{authblk}
\usepackage{multirow}
\usepackage{gensymb}

\usetikzlibrary{automata, positioning, arrows, bayesnet, matrix,shapes,chains, calc}

\usepackage[numbers]{natbib}

\usepackage[ruled, linesnumbered]{algorithm2e}

\DeclareMathOperator*{\argmax}{\arg\!\max}

\newcommand{\mmap}{\ensuremath{\mathrm{mmap}}}

\newcommand{\Pmat}{\ensuremath{\vect{P}}} 

\newcommand{\mat}[1]{\ensuremath{\boldsymbol{\mathbf{#1}}}} 

\newcommand{\const}[0]{\ensuremath{\mathrm{const}}} 
\newcommand{\tv}[0]{\ensuremath{t}} 
\newcommand{\kappat}[1]{\ensuremath{\kappa_{#1}}} 

\newcommand{\OmegaKappa}{\ensuremath{\Omega_{\kappa}}}		

\newcommand{\vect}[1]{\ensuremath{\boldsymbol{\mathbf{#1}}}}		
\newcommand{\vkappa}[0]{\ensuremath{\boldsymbol{\mathbf{\kappa}}}}		
\newcommand{\vm}[0]{\ensuremath{\boldsymbol{\mathbf{m}}}}		
\newcommand{\vr}[0]{\ensuremath{\boldsymbol{\mathbf{r}}}}		
		
\newcommand{\vd}[0]{\boldsymbol{\mathbf{d}}}		

\newcommand{\vkappatk}[2]{\ensuremath{\vkappa_{#1}^{(#2)}}}
\newcommand{\vkappak}[1]{\ensuremath{\vkappa_{#1}^{(k)}}}

\newcommand{\kk}{\ensuremath{k}}
\newcommand{\kprime}{\ensuremath{k'}}

\newcommand{\nclasses}[0]{\ensuremath{L}}		

\newcommand{\gpdf}[4]{\ensuremath{\phi_{#1}\left({#2}; {#3}, {#4} \right)}} 

\newcommand{\pdf}[1]{\ensuremath{p\!\left(#1\right)\!}} 
\newcommand{\pdff}[2]{\ensuremath{p_{#1}\left(#2\right)}} 
\newcommand{\cpdf}[2]{\ensuremath{p\!\left(\left. #1 \right| #2\right)\!}} 
\newcommand{\cpdfk}[2]{\ensuremath{p^{(k)}\!\left(\left. #1 \right| #2\right)\!}} 
\newcommand{\cpdfka}[2]{\ensuremath{p_{a}^{(k)}\left(\left. #1 \right| #2\right)}} 
\newcommand{\prior}{\ensuremath{\pdf{\vkappa}}} 

\newcommand{\posterior}{\ensuremath{\cpdf{\vkappa}{\vd}}}

\newcommand{\lik}{\ensuremath{\cpdf{\vd}{\vkappa}}}

\newcommand{\rhomgk}{\ensuremath{\rho_{m|\kappa}}}
\newcommand{\diag}{\ensuremath{\mathrm{diag}}}

\newcommand{\vardgm}{\ensuremath{\sigma_{d|m}}}
\newcommand{\mum}{\ensuremath{\vect{\mu}_{\vm}}}
\newcommand{\mumm}{\ensuremath{{\mu}_{m}}}

\newcommand{\Sm}{\ensuremath{\mat{\Sigma}_{\vm}}}

\newcommand{\mudgk}{\ensuremath{\vect{\mu}_{\vd|\vkappa}}}
\newcommand{\mumgk}{\ensuremath{\vect{\mu}_{\vm|\vkappa}}}

\newcommand{\edgm}{\ensuremath{\vect{\epsilon}_{\vd|\vm}}}

\newcommand{\emgk}{\ensuremath{\vect{\epsilon}_{\vm|\vkappa}}}

\newcommand{\Sdgm}{\ensuremath{\vect{\Sigma}_{\vd|\vm}}}
\newcommand{\Smgk}{\ensuremath{\vect{\Sigma}_{\vm|\vkappa}}}

\newcommand{\Sdgk}{\ensuremath{\vect{\Sigma}_{\vd|\vkappa}}}

\newcommand{\Srgk}{\ensuremath{\vect{\Sigma}_{\vr|\vkappa}}}

\newcommand{\sigmamgki}[1]{\ensuremath{\sigma_{m_{#1}|\kappa_{#1}}}}
\newcommand{\vmtk}[2]{\ensuremath{\vm_{#1}^{(#2)}}}

\newcommand{\vdtk}[2]{\ensuremath{\vd_{#1}^{(#2)}}}

\newcommand{\Imat}[1]{\ensuremath{\vect{I}_{#1}}}

\newcommand{\Wmat}{\ensuremath{\vect{W}}}

\newcommand{\Dmat}{\ensuremath{\vect{D}}}
\newcommand{\Amat}{\ensuremath{\vect{A}}}

\newcommand{\transpose}{\ensuremath{^\intercal}}

\newcommand{\pdfapprox}[1]{\ensuremath{p^{\star}\left(#1\right)}}

\newcommand{\rhom}{\ensuremath{\rho_{\vm}}}

\usepackage{graphicx}
\usepackage{textcomp}
\usepackage{gensymb}
\title{Bayesian inversion of convolved hidden Markov models with applications in reservoir prediction}
\author[1]{Torstein Fjeldstad\thanks{torstein.fjeldstad@ntnu.no}}
\author[1]{Henning Omre}
\affil[1]{Department of Mathematical Sciences, Norwegian University of Science and Technology, Trondheim.}

\begin{document}

\maketitle

\begin{abstract}

Efficient assessment of convolved hidden Markov models is discussed. The bottom-layer is defined as an unobservable categorical first-order Markov chain, while the middle-layer is assumed to be a Gaussian spatial variable conditional on the bottom-layer. Hence, this layer appear as a Gaussian mixture spatial variable unconditionally. We observe the top-layer as a convolution of the middle-layer with Gaussian errors. Focus is on assessment of the categorical and Gaussian mixture variables given the observations, and we operate in a Bayesian inversion framework. The model is defined to make inversion of subsurface seismic AVO data into lithology/fluid classes and to assess the associated elastic material properties. Due to the spatial coupling in the likelihood functions, evaluation of the posterior normalizing constant is computationally demanding, and brute-force, single-site updating Markov chain Monte Carlo algorithms converges far too slow to be useful. We construct two classes of approximate posterior models which we assess analytically and efficiently using the recursive Forward-Backward algorithm. These approximate posterior densities are  used as proposal densities in an independent proposal Markov chain Monte Carlo algorithm, to assess the correct posterior model. A set of synthetic realistic examples are presented. The proposed approximations provides efficient proposal densities which results in acceptance probabilities in the range 0.10-0.50 in the Markov chain Monte Carlo algorithm. A case study of lithology/fluid seismic inversion is presented. The lithology/fluid classes and the elastic material properties can be reliably predicted. 
\end{abstract}


\section{Introduction}
Inverse problems arises naturally in several fields of engineering, such as image analysis and signal processing, and constitutes a major challenge. The variables of interest are observed through an acquisition procedure together with an error term. The objective is to predict the variables of interest given the observations, being an inverse problem. We cast the problem into a Bayesian inversion framework \citep{Tarantola2005}, defined by a likelihood and a prior model.

Focus is on the class of switching state-space models where the likelihood function for the observations vary according to an unobservable categorical random process, see \cite{Schnatter2006} and references therein. Switching state-space models have numerous applications in, for example, econometrics \citep{Giordani2007}, signal processing \citep{Barembruch2009}, speech recognition \citep{Rabiner1989} and blind deconvolution \citep{Lindberg2014}.

We restrict ourselves to the subclass of convolved hidden Markov models inspired by \cite{Larsen2006}, \cite{Ulvmoen2010} and \cite{Rimstad2013}. The bottom-layer is assumed to be an unobservable categorical Markov chain. Conditional on the bottom-layer we define a middle-layer being a Gaussian spatial variable, which then appear as a Gaussian mixture spatial variable unconditionally. The top-layer, representing the convolved observations, is assumed to be Gaussian conditional on the middle-layer. Focus is on assessment of the categorical and the Gaussian mixture variables given the convolved observations, which appear as an inverse problem. 

Recursive algorithms have proven to be successful for hidden Markov models \citep{Scott2002, Reeves2004}. Indeed, they are well-suited for models where each observation depend only on current and past values of the categorical process, not future values which is the case with convolved observations. Thus, for more complex likelihood functions sample based inference by Monte Carlo sampling, such as particle filters, is often required \citep{Fearnhead2003}.

Consider a discretized vertical profile of categorical classes subsurface penetrating a reservoir unit. The categorical variable in the bottom-layer represent, for example, geological lithology/fluid classes such as gas sandstone, brine sandstone or shale. In reservoir characterisation such inverse problems are of utmost importance to predict the presence of hydrocarbons \citep{Avseth2005, Doyen2007, Gunning2007}. The middle-layer, being a Gaussian mixture spatial variable, represents, for example, the elastic material properties of the reservoir. \cite{Buland2003} proposed a linearized Bayesian inversion technique for assessment of continuous-valued properties subsurface, however they did not account for the different effects of the lithology/fluid classes. \cite{Grana2010} proposed a Gaussian mixture density prior model for seismic velocities, but their model did not included spatial dependence in the middle-layer. \cite{Connolly2016} proposed an approximate sampling technique based on approximate Bayesian computation to assess lithology/fluid classes and the continuous-valued properties.

We consider an extension of the convolved hidden Markov model defined and evaluated in \cite{Larsen2006} and \cite{Rimstad2013}. We assume a Markov chain prior model for the bottom-layer. In our model the middle-layer is  Gaussian spatial variable conditional on the bottom-layer defined by class-dependent expectations and variances, and a spatial correlation function. The spatial correlation is defined to enforce spatial continuity in the middle-layer, a property observed in subsurface seismic velocities. Unconditionally the middle-layer appear as a Gaussian mixture spatial model, extending the traditional Gaussian prior assumption for seismic velocities \citep{Buland2003}. The top-layer contains convolved observations, which appear unconditionally as a Gaussian mixture random variable dependent on past, current and future values of the two lower layers. Assessment of the categorical variable given the observations, which constitutes a challenging categorical inverse problem, is discussed. Moreover, we discuss simulation and prediction of the middle-layer Gaussian mixture variable. 

The convolution and spatial coupling in the likelihood model prevents straightforward use of recursive algorithms since the posterior model can not be written on factorial form \citep{Reeves2004}. \cite{Larsen2006}, \cite{Hammer2011} and \cite{Rimstad2013} proposed various approximate posterior models on lower order factorial form which are computationally feasible. For the extended model defined in our study, we present two classes of likelihood approximations and demonstrate that their posterior models can be written on factorial form. Hence, the approximate posterior models can be efficiently assessed by the recursive Forward-Backward algorithm \citep{Baum1970}. The correct posterior model is assessed by an independent proposal Markov chain Monte Carlo (MCMC) Metropolis-Hastings (MH) algorithm, with the approximate posterior model as proposal density. In general, independent proposal densities results in a poor rate of convergence. \cite{Rimstad2013} verified in a simulation study that it is possible to obtain satisfactory acceptance rates with their proposed approximate posterior models. 

Denote by $\pdf{\cdot}$ a generic probability distribution for both categorical and continuous random variables. The probability density function (pdf) for a $n$-dimensional Gaussian random variable \vect{x}\ having mean \vect{\mu}\ and covariance matrix \mat{\Sigma} is denoted by \gpdf{n}{\vect{x}}{\vect{\mu}}{\mat{\Sigma}}. We refer to a likelihood model that is linear in the conditioning variable with additive Gaussian errors as a Gauss-linear likelihood model. Let $\mat{I}_n$ be the $n \times n$ identity matrix.

In Section~\ref{sec:Model} we define the convolved hidden Markov model. Section~\ref{sec:Likelihood} contains the definition of the proposed likelihood approximations, and a MCMC algorithm to assess the correct posterior model is presented in Section~\ref{sec:posterior}. A simulation study is included in Section~\ref{sec:Synthetic}, where we consider various model parameters to empirically evaluate the overall performance of the two proposed likelihood approximations. The synthetic simulation models are chosen to be comparable to the ones in \cite{Rimstad2013}. A seismic inversion case study based on real data is included in Section~\ref{sec:RCS}.
%
\section{Model specification}
\label{sec:Model}
Consider an unobservable categorical variable $\vkappa = \left( \kappat{1}, \ldots, \kappat{n} \right)\transpose$ on a discretized top-down regular grid $\mathcal{L} : \{1, \ldots, n\}$, along a vertical profile, representing for example lithology/fluid classes of the subsurface. Label the variable $\kappat{\tv} \in \OmegaKappa=\{1,\ldots, \nclasses\}$ for $\tv \in \mathcal{L}$. A continuous valued variable $\vd = \left( d_1, \ldots, d_{n}\right)\transpose \in \mathbb{R}^n$ is observed and represent, for example, seismic data. The main objective is to predict \vkappa\ given \vd\ with the associated uncertainty, being a categorical inverse problem. We operate in a Bayesian framework, and the posterior density for the categorical variable given the observations is defined by Bayes' theorem:
\begin{equation}
	\label{eq:posterior}
	\posterior = \const \times \lik \prior,
\end{equation}  
where \const\ is the normalizing constant, \lik\ is the likelihood function, and \prior\ is the prior model. The solution of the categorial inverse problem in Eq.~\eqref{eq:posterior} is a computationally challenging problem, and analytical assessment is only feasible for very particular models. We introduce an additional continuous valued variable $\vm=\left( m_1,\ldots,m_{n}\right)\transpose \in \mathbb{R}^{n}$ separating \vd\ and \vkappa. That is, we assume \vd\ and \vkappa\ to be conditionally independent given \vm. The variable \vm\ represents, for example, the logarithm of the seismic velocity. We define the gross likelihood model \citep{Larsen2006} as:
\begin{equation}
	\label{eq:lik}
	\lik = \int \cpdf{\vd}{\vm}\cpdf{\vm}{\vkappa}\dif{\vm},
\end{equation}
where \cpdf{\vd}{\vm} and \cpdf{\vm}{\vkappa} are respectively the acquisition and response likelihood functions. The variable \vm\ is marginalized out in \cpdf{\vd}{\vkappa}, but its uncertainty propagates into \lik. In general, integrating out \vm\ in Eq.~\eqref{eq:lik} is a challenging problem caused by its dimension, and analytical expressions need not exist.

\subsection{Prior model}
We define a Markov chain prior model to represent vertical dependency in the categorical variable of interest.
Let the categorical variable \vkappa\ be defined by a $\kk$-th order stationary Markov chain,
\begin{equation}
	\label{eq:prior}
	\begin{aligned}
	\prior & = \pdff{s}{\kappa_1, \ldots, \kappa_{\kk}}\!\times\! \prod_{\tv = \kk+1}^{n}\! \cpdf{\kappa_{\tv}}{\kappa_{\tv{-1}},\ldots, \kappa_{\tv-\kk}} \\
	& =  \pdff{s}{\vkappatk{\tv}{\kk}} \times \prod_{\tv = \kk+1}^n \cpdf{\vkappatk{\tv}{\kk}}{\vkappatk{\tv-1}{\kk}}
	\end{aligned},
\end{equation}
with $\vkappatk{\tv}{\kk} = \left( \kappa_{\tv-\kk+1},\ldots, \kappa_{\tv}\right)\transpose$ for $\tv= \kk, \ldots, n$. The latter equality is justified by a trivial extension of the sample space since the set $\left( \kappa_{\tv-\kk+1}, \ldots, \kappa_{\tv-1}\right)$ is a subset of the conditioning set. The corresponding transition $\left( \nclasses^{\kk} \times \nclasses^{\kk}\right)$-matrix \Pmat\ is time-independent and contain at most $\nclasses^{\kk+1}$ non-zero elements, due to the trivial extension. Let $ \pdff{s}{\vkappatk{\tv}{\kk}}$ be the corresponding stationary distribution. Indeed in this case the spatial Markov property is $\cpdf{\vkappa_t}{\vkappa_{-t}} = \cpdf{\kappa_t}{\vkappatk{t-1}{k}, \vkappatk{t+k}{k}}$, and the prior model is a $k$-th order Markov random field \citep{Besag1974}.


\subsection{Likelihood model}
\label{sec:likelihood}

We define the response model as follows:
\label{sec:response}
\begin{equation}
	[\vm|\vkappa] = \mumgk + \emgk,
\end{equation}
where $\mumgk = \left( \mu_{m_1|\kappa_1},\ldots, \mu_{m_{n}|\kappa_{n}}\right)\transpose \in \mathbb{R}^{n}$ is the vector with pointwise conditional means given \vkappa, and $\emgk\in\mathbb{R}^{n}$ is a centered Gaussian process having covariance matrix $\Smgk\in\mathbb{R}^{n\times n}$. 
The response likelihood is therefore Gaussian:
\begin{equation}
\label{eq:response-lik}
	\cpdf{\vm}{\vkappa} = \gpdf{n}{\vm}{\mumgk}{\Smgk}.
\end{equation}
We decompose the covariance matrix \Smgk\ as follows:
\begin{equation}
	\label{eq:cov-response}
	\Smgk = \Smgk^{\sigma}\Sm^{\rho}\Smgk^{\sigma},
\end{equation}
where $\Smgk^{\sigma} =  \diag\left(\sigmamgki{1}, \ldots,\sigmamgki{n}  \right) \in \mathbb{R}^{n \times n}$ is a diagonal matrix with the conditional standard deviations. The correlation matrix $\Sm^{\rho}\in \mathbb{R}^{n\times n}$ is defined from a spatial correlation function $\rhom(h)$, where $h = | {\tv}-{s}|$ for all combinations of $t,s = 1,\ldots, n$. This entails that the response process \vm\ is constructed by a normalized Gaussian process which is scaled and shifted dependent on the current categorical variable for each $t = 1, \ldots, n$. For each $t = 1,\ldots, n$, the Gaussian process representing the current categorical variable is assigned. This response process can be extended to have a set of $L$ correlated Gaussian processes, with separate expectations, variances and spatial correlation functions - one for each of the $\nclasses$ classes of the categorical variable. The inversion methodology defined in the following sections work also for this case, but the notation will be more complex. We present only the simpler parametrization in Eq.~\eqref{eq:response-lik}-\eqref{eq:cov-response} to ease readability.

The marginal pdf of \vm:
\begin{equation}
	\label{eq:dist-m}
	\pdf{\vm} = \sum_{\vkappa}\cpdf{\vm}{\vkappa}\pdf{\vkappa},
\end{equation}
is a $n$-dimensional Gaussian mixture pdf with at most $L^n$ unique modes. Even for short profiles, evaluation of \pdf{\vm}\ is unfeasible since it requires evaluation of $\nclasses^n$ different configurations of \vkappa. Also, for $t = 1, \ldots, n$, the unconditional pdfs
\begin{equation}
	\label{eq:pdf-mt}
	\pdf{m_{\tv}} = \sum_{\kappa_{\tv}}\gpdf{1}{m_{\tv}}{\mu_{m_{\tv}|\kappa_{\tv}}}{\sigma_{m_{\tv}|\kappa_{\tv}}^2}\pdf{\kappa_{\tv}},
\end{equation}
are univariate Gaussian mixture pdfs with at most $L$ unique modes.

We consider a convolved data acquisition procedure represented by the convolution matrix $\Wmat\in \mathbb{R}^{n \times n}$, representing time-independent convolutions. Hence, each datum $d_{\tv}$ appear as a weighted sum of neighboring elements of the response variable \vm. Consider the following linear acquisition model:
\begin{equation}
	[\vd|\vm] = \Wmat\vm + \edgm,
\end{equation}
where $\edgm\in \mathbb{R}^{n}$ is a centered Gaussian process with covariance matrix $\Sdgm\in\mathbb{R}^{n\times n}$. Thus, the acquisition likelihood model:
\begin{equation}
	\label{eq:acq-lik}
\cpdf{\vd}{\vm}=\gpdf{n}{\vd}{\Wmat\vm}{\Sdgm}
\end{equation} is Gaussian with mean vector $\Wmat\vm$ and covariance matrix \Sdgm. Indeed, the proposed acquisition likelihood is straightforward to generalize to cases where \vd\ and \vm\ are of different dimension.

Since both the response and acquisition likelihood models are assumed to be Gaussian, the challenging gross likelihood model in Eq.~\eqref{eq:lik} is also Gaussian:
\begin{equation}
\begin{aligned}
	\label{eq:lik-gross}
	\lik & = \gpdf{n}{\vd}{\Wmat\mumgk}{\Wmat\Smgk\Wmat\transpose + \Sdgm} \\ &= \gpdf{n}{\vd}{\mudgk}{\Sdgk}
	\end{aligned}.
\end{equation}Given \vkappa, each datum $d_{\tv}$ has an expectation as a weighted sum of the conditional mean vector \mumgk. Note that the explicit dependence on \vkappa\ in the covariance matrix requires the covariance matrix to be computed for each unique \vkappa, thus evaluation of the likelihood is computationally expensive for a set of unique \vkappa.

%

\subsection{Posterior model}

\begin{figure*}[ht!]
	\centering
	\begin{tikzpicture}[->,>=stealth', auto,semithick,node distance=3cm]
		\tikzstyle{every state}=[fill=white,draw=black,thick,text=black,scale=0.5, minimum size=1.5cm, inner sep=1]
		\node[state]    	(pi1)						{$\kappa_1$};
		\node[draw = none] (pi15)[right of = pi1, node distance = 1.5cm]	{$\hdots$};
		\node[state]    	(pi2)[right of = pi15]		{$\kappa_{\tv-1}$};
		\node[state] 	(pi3)[right of = pi2]		{$\kappa_{\tv}$};
		\node[state]		(pi4)[right of = pi3]		{$\kappa_{\tv+1}$};
		\node[draw = none] (pi45)[right of = pi4, node distance = 1.5cm]	{$\hdots$};
		\node[state]		(pi5)[right of = pi45]		{$\kappa_{n}$};

		\node[state]		(r1)[above of = pi1]		{$m_1$};
		\node[draw = none] (r15)[right of = r1, node distance = 1.5cm]	{$\hdots$};
		\node[state]		(r2)[above of = pi2]			{$m_{\tv-1}$};
		\node[state]		(r3)[above of = pi3]		{$m_{\tv}$};
		\node[state]		(r4)[above of = pi4]			{$m_{t-1}$};
		\node[draw = none] (r45)[right of = r4, node distance = 1.5cm]	{$\hdots$};
		\node[state]		(r5)[above of = pi5]			{$m_{n}$};

		\node[state]		(d1)[above of = r1, node distance = 5cm]			{$d_1$};
		\node[state]		(d2)[above of = r2, node distance = 5cm]			{$d_{\tv-1}$};
		\node[state]	(d3)[above of = r3, node distance = 5cm]			{$d_{\tv}$};
		\node[state]		(d4)[above of = r4, node distance = 5cm]			{$d_{\tv+1}$};
		\node[state]		(d5)[above of = r5, node distance = 5cm]			{$d_{n}$};
		\node[draw = none] (d15)[right of = d1, node distance = 1.5cm]	{$\hdots$};
		\node[draw = none] (d45)[right of = d4, node distance = 1.5cm]	{$\hdots$};

\edge {pi1} {pi15}
\edge{pi15} {pi2}
		\edge {pi2} {pi3}
		\edge {pi3} {pi4}
						\edge {pi4} {pi45}
			\edge {pi45} {pi5}
			
			\edge {pi15} {pi1}
\edge{pi2} {pi15}
		\edge {pi3} {pi2}
		\edge {pi4} {pi3}
						\edge {pi45} {pi4}
			\edge {pi5} {pi45}

		\edge {pi1} {r1}
		\edge {pi2} {r2}
		\edge {pi4} {r4}
		\edge {pi5}	{r5}
		\edge {pi1} {r15}
		\edge {pi2} {r15}
		\edge {pi4} {r45}
		\edge {pi5}	{r45}
		\edge {pi2} {r3}
		\edge {pi4} {r3}

		\edge {pi3} {r3}
		\edge {pi3} {r2}
		\edge {pi3} {r4}
		\edge {r1} 	{d1}
		
		\edge {r2}	{d2}
		\edge {r4}	{d4}
		\edge {r5}	{d5}

		\edge {r1} 	{d1}
		\edge {r1}	{d15}
		\edge {r2}	{d15}
		\edge {r2} 	{d2}
		\edge {r2} 	{d3}
		\edge {r2} 	{d4}
		\edge {r2} 	{d45}
		\edge {r3}	{d15}
		\edge {r3}	{d2}
		\edge {r3} {d3}
		\edge {r3}	{d4}
		\edge {r3}	{d45}
		\edge {r4}	{d15}
		\edge {r4}	{d2}
		\edge {r4}	{d45}
		\edge {r4}	{d3}
		\edge {r4}	{d4}
		\edge {r5}	{d45}
		\edge {r15} {d1}
		\edge {r15} {d2}
		\edge {r15} {d3}
		\edge {r15} {d4}
		\edge {r45} {d5}
		\edge {r45} {d2}
		\edge {r45} {d3}
		\edge {r45} {d4}
				
\edge {r1} {r15}
\edge{r15} {r2}
		\edge {r2} {r3}
		\edge {r3} {r4}
						\edge {r4} {r45}
			\edge {r45} {r5}
			
			\edge {r15} {r1}
\edge{r2} {r15}
		\edge {r3} {r2}
		\edge {r4} {r3}
						\edge {r45} {r4}
			\edge {r5} {r45}

	\end{tikzpicture}
	\caption{Convolved hidden Markov model with \prior\ being a first order Markov chain, $\cpdf{\vm}{\vkappa}$ having a first order exponential covariance function, and \Wmat\ having finite support.}
		\label{fig:convolved-model2}
\end{figure*}
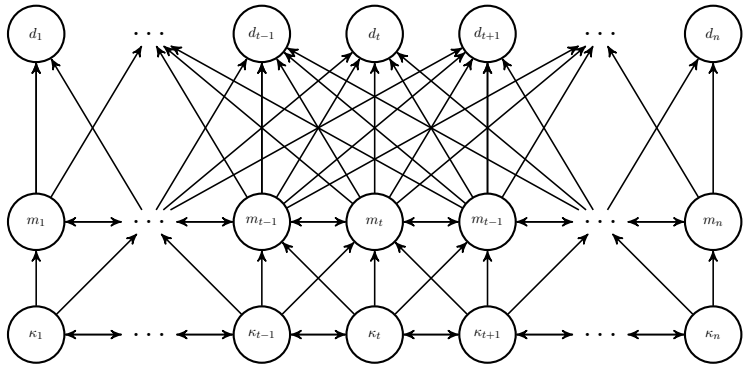

The model defined in the previous section may be represented by a graph. If we assume that the spatial correlation function $\rhom(h)$ is on first order exponential form, the convolution kernel defining \Wmat\ has finite support and the prior model is a first order Markov chain, the graph takes a simple form (Fig.~\ref{fig:convolved-model2}). Indeed, each observation $d_t$ is seen to be dependent on a large subset of \vkappa.

The general posterior model is
\begin{equation}
	\posterior = \const \times \int \cpdf{\vd}{\vm}\cpdf{\vm}{\vkappa}\dif{\vm}\times \prior,
\end{equation}
where
\begin{equation}
	\const = \left[\sum_{\vkappa } \int \cpdf{\vd}{\vm}\cpdf{\vm}{\vkappa}\dif{\vm} \times \prior \right]^{-1}
\end{equation}
is computationally challenging since it requires summing over $\nclasses^{n}$ elements. The posterior distribution is the ultimate solution to the Bayesian inversion problem, but it is often represented by a set of independent realizations $\vkappa^1, \ldots, \vkappa^B$ from $\cpdf{\vkappa}{\vd}$. Characteristics such as marginal probability profiles MPR$^\kappa$ for $\kappa \in \Omega_{\kappa}$:
\begin{equation}
	\vect{p}_\mathrm{MPR}^{\kappa} = \left\{\cpdf{\kappa_t = \kappa}{\vd}  ; \quad t = 1,\ldots, n\right\},
\end{equation}
maximum posterior (MAP) predictor
\begin{equation}
	\vkappa_{\mathrm{MAP}} = \argmax_{\vkappa} \cpdf{\vkappa}{\vd},
\end{equation}
or alternatively the marginal maximum posterior (MMAP) predictor
\begin{equation}
	\vkappa_{\mathrm{MMAP}} = \left\{ \argmax_{\kappa} \cpdf{\kappa_t = \kappa}{\vd}; \quad t = 1,\ldots, n\right\}
\end{equation}
are often used to describe the distribution. Since the MMAP predictor is only a sequence of pointwise maximums, it need not necessarily equal the computationally unfeasible global MAP predictor. Indeed, transitions having zero-probability in the prior model may occur in the MMAP since the latter is only a pointwise property.

It can be shown that the posterior model \cpdf{\vm}{\vd}\ is also a Gaussian mixture density
\begin{equation}
	\cpdf{\vm}{\vd} = \sum_{\vkappa} \gpdf{n}{\vm}{\vect{\mu}_{\vm|\vd, \vkappa}}{\mat{\Sigma}_{\vm|\vd, \vkappa}}\cpdf{\vkappa}{\vd},
\end{equation} where the model parameters $\vect{\mu}_{\vm|\vd, \vkappa}$ and ${\mat{\Sigma}_{\vm|\vd, \vkappa}}$ are as given in \cite{Grana2016}. That is, \pdf{\vm} is a conjugate prior model with respect to the Gaussian likelihood function \cpdf{\vd}{\vm}. We define the following MMAP predictor:
\begin{equation}
	\label{eq:posterior-mt}
	\hat{m}_t \!=  \!\argmax_{m} \left\{ \sum_{\vkappa}\gpdf{1}{m}{\mu_{m|\vd, \vkappa}}{\sigma_{m|\vd, \vkappa}^2}\posterior \right\} 
\end{equation}for $t = 1,\ldots, n$, being $n$ univariate optimizations.  Here, $\mu_{m|\vd, \vkappa}$ is the $t$-th element in $\vect{\mu}_{\vm|\vd, \vkappa}$ and $\sigma_{m|\vd, \vkappa}^2$ is the corresponding $t$-th diagonal element in $\mat{\Sigma}_{\vm|\vd, \vkappa}$. Posterior $100 \times (1-\alpha)$ \% prediction intervals are obtained similarly. 

In a generic convolved hidden Markov model there are severe couplings, mostly due to the convolution operator \Wmat\ and spatial dependence. Assessment of the posterior model by brute-force single-site proposal MCMC algorithms is not feasible. We choose to approximate the likelihood model such that we obtain an approximate posterior model which is analytically tractable. This approximate posterior model is afterwards used as a proposal distribution in an independent proposal MCMC MH algorithm to assess the correct posterior model. For a simpler model \cite{Rimstad2013} obtained satisfactory acceptance rates in a simulation study.

%
\section{Likelihood approximations}
\label{sec:Likelihood}
We present two likelihood approximations of the gross likelihood, inspired by \cite{Rimstad2013}, to obtain approximate posterior models on low-order factorial form. Such models are efficiently assessed by recursive algorithms. We consider only approximations of the likelihood function \cpdf{\vd}{\vkappa} denoted by $\cpdfk{\vd}{\vkappa}$ for $k = 1,2,\ldots$, since the spatial correlation in the response likelihood and the convolution acquisition likelihood contribute with the major spatial couplings in the model. Recall that a likelihood function, contrary to a probability density, need not be normalized, hence the former is scale invariant.

Define $\vmtk{\tv}{\kk} = \left( m_{\tv-\kk+1},\ldots, m_{\tv}\right)\transpose$ for $\tv = \kk,\ldots, n$, similarly as \vkappatk{\tv}{\kk}. From the definition of the response likelihood function it follows that:
\begin{equation}
	\label{eq:mtk-given-kappatk}
	\cpdf{\vmtk{\tv}{\kk}}{\vkappa} = \cpdf{\vmtk{\tv}{\kk}}{\vkappatk{\tv}{\kk}}.
\end{equation}
The proposed approximations are based on, respectively, a na\"ive truncation of the likelihood function and a Gaussian approximation of the Gaussian mixture pdf \pdf{\vm}.

The proposed likelihood approximations should be such that {$\|\cpdfk{\vd}{\vkappa}-\cpdf{\vd}{\vkappa} \|$} is small with respect to some measure, and decrease for increasing \kk. In practice, we have to accepted a low-order approximation since we have to re-run the approximation for different model parameter values.

\subsection{Truncation approximation}
\label{sec:truncation}
First we consider a na\"{\i}ve approximation, inspired by Approximation 1 in \cite{Rimstad2013}, by truncating the marginal densities. For simplicity we assume $\Sdgm = \vardgm^2 \Imat{n}$, since coloured errors in \Sdgk\ appear as results of the convolution. Thus, the acquisition likelihood can be written on factorial form:
\begin{equation}
	\cpdf{\vd}{\vm} = \prod_{\tv=1}^{n} \cpdf{d_{\tv}}{\vm}.
\end{equation}

Define the $k$-band truncated matrix $\Wmat^k$ by truncating every element in \Wmat\ more than $\kprime$ from the diagonal equal to zero, where $\kprime$ is such that $\kk = 2\kprime+1$ for $\kprime = 0,1,\ldots$. Thus, it follows that the $\kk$-th order marginal acquisition likelihood for $t = \kprime+2,\ldots, n-\kprime-1$ is given as:
\begin{equation}
	\label{eq:pdf-truncation-dt-given-m}
	\cpdfk{d_{\tv}}{\vm} = \gpdf{1}{d_{\tv}}{ \vect{w}_{\tv}^{\kk}\vm}{\vardgm^2},
\end{equation}
where $\vect{w}_{\tv}^{\kk}$ is the $t$-th row in $\Wmat^k$. Combining Eq.~\eqref{eq:mtk-given-kappatk} and Eq.~\eqref{eq:pdf-truncation-dt-given-m} we obtain
\begin{equation}
	\label{eq:pdf-truncation-dt-given-kappatk}
	\cpdfk{d_{\tv}}{\vkappak{{\tv}+\kprime}} \! = \!\int \!\cpdfk{d_{\tv},\vmtk{\tv+\kprime}{\kk}}{\vkappatk{\tv+\kprime}{\kk}} \dif{\vmtk{\tv+\kprime}{\kk}},
\end{equation}
which are Gaussian pdfs for $\tv= \kprime+2, \ldots, n-\kprime-1$. We define the $\kk$-th order truncation likelihood approximation as:

\begin{equation}
	\label{eq:truncation}
	\begin{aligned}
	\cpdfk{\vd}{\vkappa} & = \cpdfk{\vdtk{\kprime+1}{\kprime+1}}{\vkappatk{\kk}{\kk}} \\
	& \quad \times \prod_{\tv=\kprime+2}^{n - \kprime-1} \cpdfk{d_t}{ \vkappatk{\tv+\kprime}{\kk}}\\ & \quad \times \cpdfk{\vdtk{n}{\kprime+1}}{\vkappatk{n}{\kk}} \end{aligned},
\end{equation}where first and last factor are boundary correction terms. Note that each $\vkappatk{t}{k}$ is only dependent on a small subset of the observations. Indeed, if \Wmat\ and \Sdgk\ are diagonal matrices the truncation approximation of order one is exact, and the model correspond to a standard hidden Markov model. Note that we do not require \Sdgk\ to be independent of \vkappa.
The definition above is based on $\Wmat\in \mathbb{R}^{n\times n}$ representing a time-invariant convolution, but it can be extended to an arbitrary linear operator \Wmat\ by $\kk$-truncating each row such that it covers as much weight as possible, and by extracting the corresponding subset of \vkappa.

\subsection{Projection approximation}
The second proposed likelihood approximation, which we denote the projection approximation, is based on a Gaussian approximation of the Gaussian mixture pdf \pdf{\vm}. Let \pdfapprox{\vm} be the Gaussian approximation with mean

\begin{equation}
		\left[\mum\right]_{\tv}  =  \sum_{\kappa_{\tv}}\mu_{m_{\tv}|\kappa_{\tv}}\pdff{s}{\kappa_{\tv}} = \mu_m;  \qquad  t = 1, \ldots, n,
\end{equation}
and covariance
\begin{equation}
\begin{aligned}
		\left[\Sm\right]_{\tv, s} \! & = \! \sum_{\kappa_{\tv}}\sum_{\kappa_{s}}\pdf{\kappa_{\tv},\kappa_{s}}\left\{\sigma_{m_{\tv}|\kappa_{\tv}}\sigma_{m_{s}|\kappa_{s}}\rhomgk({t-s})\right. \\
		& \left.\quad +\left(\mu_{m_{\tv}|\kappa_{\tv}}-\mumm \right)\left( \mu_{m_{s}|\kappa_{s}}-\mumm\right)\right\}
	\end{aligned}
\end{equation}
for $\tv = 1,\ldots, n, s = 1,\ldots, n$. These expressions are obtained analytically using the laws of total expectation and covariance.

The joint approximate distribution $\pdfapprox{\vd, \vm} = \cpdf{\vd}{\vm}\pdfapprox{\vm}$ is a Gaussian pdf, thus also the marginal distributions $\pdfapprox{ \vd, \vmtk{\tv}{\kk}}$ for $\tv = \kk,\ldots, n$ are Gaussian pdfs. By conditioning, we define the $\kk$-th order approximate acquisition likelihood model $\cpdfk{\vd}{\vmtk{\tv}{\kk}}$ for $\tv = \kk, \ldots, n$, which are Gaussian pdfs. Combining the results above with Eq.~\eqref{eq:mtk-given-kappatk} it can be verified that
\begin{equation}
\label{eq:projection-d-given-kappatk}
\cpdfk{\vd}{\vkappak{{\tv}}}  = \int \cpdfk{\vd,\vmtk{\tv}{\kk}}{\vkappatk{\tv}{\kk}}  \dif{\vmtk{\tv}{\kk}}
\end{equation}are Gaussian pdfs for $\tv = \kk, \ldots, n$. We note that the marginalization requires evaluation of a $n$-dimensional Gaussian pdf, $\cpdfk{\vd}{\vmtk{\tv}{\kk}}$, which is computationally expensive for large $n$. Indeed, by Bayes' theorem, we have that 
\begin{equation}
	\cpdfk{\vd}{\vmtk{\tv}{\kk}} = \frac{\cpdfk{\vmtk{\tv}{\kk}}{\vd}p\left(\vd\right) }{p^{(k)}\left( \vmtk{\tv}{\kk}\right)}.
\end{equation}
Thus, we rephrase Eq.~\eqref{eq:projection-d-given-kappatk} as:
\begin{equation}
	\begin{aligned}
	\cpdfk{\vd}{\vkappak{{\tv}}} & \propto \int  \frac{\cpdfk{\vmtk{\tv}{\kk}}{\vd} }{p^{(k)}\left( \vmtk{\tv}{\kk}\right)} \cpdf{\vmtk{\tv}{\kk}}{\vkappatk{\tv}{\kk}}  \dif{\vmtk{\tv}{\kk}} \\
	& \propto  \int  \frac{\cpdfk{\vmtk{\tv}{\kk}}{\vd} }{p^{(k)}\left( \vmtk{\tv}{\kk}\right)} \cpdf{\vmtk{\tv}{\kk}}{\vkappatk{\tv}{\kk}}  \dif{\vmtk{\tv}{\kk}}
	\end{aligned},
\end{equation}
where the densities to be evaluated are of dimension $\kk << n$.

We define the $\kk$-th order projection approximation as follows:
\begin{equation}
	\label{eq:projection}
	\begin{aligned}
	\cpdfk{\vd}{\vkappa} & = \left\{\cpdfk{\vd}{\vkappatk{\kk}{\kk}}\prod_{i=1}^{\kk-1} \cpdfk{\vd}{\vkappatk{\kk-i}{\kk-i}} \right\}^{1/\kk} \\ & \quad \times\left\{  \prod_{\tv = \kk+1}^{n-1} \cpdfk{\vd}{\vkappatk{\tv}{\kk}}\right\}^{1/\kk}  \\
	& \quad \times \left\{\cpdfk{\vd}{\vkappatk{n}{\kk}} \prod_{i=1}^{\kk-1} \cpdfk{\vd}{\vkappatk{n}{\kk-i}}d\right\}^{1/\kk}
	\end{aligned}, 
\end{equation}where the $\kk$-root ensures that each datum is used exactly once, and the first and last factors are boundary corrections. 

Note that the projection approximation is straightforward to generalize to an arbitrary linear operator \Wmat. Contrary to the truncation approximation, the full set of observations \vd\ is used for each $\vkappatk{\tv}{\kk}$ in the projection approximation. 

\section{Assessment of posterior model}
\label{sec:posterior}
Assessment of the correct posterior model $\cpdf{\vkappa}{\vd}$ by brute-force single-site simulation is unfeasible due to spatial and convolutional coupling in the observations, and possible prior ordering constraints on the categorical variable. We define an independent proposal MCMC MH algorithm based on an approximate posterior model $\cpdfk{\vkappa}{\vd}$, where the latter is exactly accessed by a recursive algorithm. 

Both the truncation and projection based approximate likelihoods functions can be phrased on factorial form
\begin{equation}
	\begin{aligned}
	\cpdfk{\vd}{\vkappa} & =   \cpdfka{\vd}{\vkappatk{\kk}{\kk}} \times \prod_{\tv=\kk+1}^{n-1}\cpdfka{\vd}{\vkappatk{\tv}{\kk}}\\
	& \quad \times \cpdfka{\vd}{\vkappatk{n}{\kk}}\end{aligned},
\end{equation}
where the subscript $a$ denotes the chosen approximation type. Thus, their approximate posterior densities can be phrased as follows:
\begin{equation}
\begin{aligned}
	\label{eq:posterior-approx}
	\cpdfk{\vkappa}{\vd} & \propto \times \cpdfka{\vd}{\vkappatk{\kk}{\kk}}\pdf{\vkappatk{\kk}{\kk}} \\
	& \quad \times \prod_{\tv = \kk+1}^{n}  \cpdfka{\vd}{\vkappatk{\tv}{\kk}}\cpdfka{\vkappatk{\tv}{\kk}}{\vkappatk{\tv-1}{\kk}} \\
	& \propto \cpdfka{\vkappatk{\kk}{\kk}}{\vd} \times  \prod_{\tv = \kk + 1}^n \cpdfka{\vkappatk{\tv}{\kk}}{\vkappatk{\tv-1}{\kk}, \vd}
	\end{aligned},
\end{equation}which is a \kk-th order non-homogeneous Markov chain. Such factorisable posterior models are exactly and efficiently assessed by the recursive Forward-Backward algorithm in $\mathcal{O}\left((n-\kk+1)\times \nclasses^{\kk+1} \right)$ operations \citep{Reeves2004}. For a given approximation order $k$, the marginal probability (MPR) profiles for the approximate posterior model is denoted aMPR$_k^\kappa$ for $\kappa \in \Omega_\kappa$ and $k \in \mathbb{Z}^+$. Correspondingly we denote the maximum posterior (MAP) and marginal maximum posterior (MMAP) predictors, respectively, aMAP$_k$ and aMMAP$_k$. These characteristics are exactly calculated for the approximate posterior model. Assessment of aMAP$_k$ requires the use of the recursive Viterbi algorithm \citep{Viterbi1967}. We refer to \cite{Lindberg2015} for a discussion of model parameter estimation in a convolved hidden Markov model.


The correct posterior model \posterior\ is assessed by an independent proposal MCMC MH algorithm \citep{Robert2005} using \cpdfk{\vkappa}{\vd} as proposal distribution. At each iteration, the acceptance probability is given as
\begin{equation}
	\label{eq:acceptance-rate}
	\alpha\left(\vkappa \to \tilde{\vkappa}\right) = \min \left\{1, \frac{\cpdf{\tilde{\vkappa}}{\vd}}{\cpdf{\vkappa}{\vd}}\times \frac{\cpdfk{\vkappa}{\vd}}{\cpdfk{\tilde{\vkappa}}{\vd}} \right\},
\end{equation}
where the troublesome normalizing constant in Eq.~\eqref{eq:posterior} cancels. After burn-in we generate $B$ realizations $\vkappa^b$ for $b = 1, \ldots, B$ from the correct posterior model $\cpdf{\vkappa}{\vd}$. To empirically quantify the quality of the proposed approximations we consider the similarity measure defined in \cite{Rimstad2013}:
\begin{equation}
	\alpha^{(k)} = \mathrm{E}_{\vkappa, \tilde{\vkappa}}\left\{ \min \left\{1, \frac{\cpdf{\tilde{\vkappa}}{\vd}}{\cpdf{\vkappa}{\vd}}\times \frac{\cpdfk{\vkappa}{\vd}}{\cpdfk{\tilde{\vkappa}}{\vd}} \right\}\right\},
\end{equation}
being the average acceptance rate in the MCMC algorithm. We define $\alpha_{\tv}^{(\kk)}$ and $\alpha_{p}^{(\kk)}$ to be the acceptance rates based on, respectively, the \kk -th order truncation and projection approximation. Acceptance rates close to unity entails that \cpdfk{\vkappa}{\vd} is close to \cpdf{\vkappa}{\vd}, which is as desired. We define performance measures $\beta_t^{(k)} = \alpha_{\tv}^{(\kk)} /L^{k-1}$ and $\beta_p^{(k)} = \alpha_{p}^{(\kk)} /L^{k-1}$ to compare the effect of the approximation order as a function of computational demands.

We estimate the marginal probabilities profiles MPR$^\kappa$ for the correct posterior model for each $\kappa \in \Omega_\kappa$ by: 
\begin{equation}
	\hat{\vect{p}}_{\mathrm{MPR}}^\kappa = \left\{ \hat{p}_t^\kappa = \frac{1}{B}\sum_{b = 1}^B \mathrm{I}\left( \kappa_t^b = \kappa\right); \quad t = 1,\ldots, n\right\}, 
\end{equation}
and similarly the marginal MAP predictor by
\begin{equation}
	\label{eq:MMAP-kappa}
	\hat{\vkappa}_{\mmap} = \left\{ \hat{\kappa}_{\tv} = \argmax_{\kappa} \ \hat{p}_t^\kappa; \quad \tv = 1,\ldots, n \right\}.
\end{equation}


Evaluating the correct posterior density \cpdf{\vkappa}{\vd} is computationally expensive since each iteration requires evaluation of a $n$-dimensional Gaussian pdf $\cpdf{\vd}{\vkappa}$. In practice, this limits the number of realizations to be generated in reasonable time. However, we note that the MCMC-step is essentially independent of the approximation order $k$; that is, for a fixed computational budget it can be beneficiary to generate few realizations from a high order approximation than many realizations from a low order approximation.


Zero-transitions in the prior matrix \Pmat\ enforces zero-transitions in the posterior, thus the $\kk$-th order approximation can be obtained at a cost smaller than the theoretical $\mathcal{O}\left((n-\kk+1)\times \nclasses^{\kk+1} \right)$ for a full matrix \Pmat.


%
\section{Synthetic example}
\label{sec:Synthetic}
We empirically evaluate the proposed likelihood approximations for various order $\kk$ in a synthetic example. The synthetic example is defined from a Base case and six deviating cases, having different model parameters. We avoid the limiting cases where the covariance matrix in the gross likelihood tends toward a diagonal matrix, which is the well-known standard hidden Markov model. That is, we consider only deviating cases where the number of non-zero entries in the covariance matrix $\Sdgk$ is at least as high as for the Base case. For the standard hidden Markov model the truncation approximation is exact for all $k$, and the projection approximation is empirically verified to have an acceptance rate close to unity for all \kk. We claim that the chosen deviating cases are the most challenging ones, since they appear with strong spatial coupling in the likelihood function.

\begin{figure*}[ht!]
\centering
	\includegraphics[width = \textwidth]{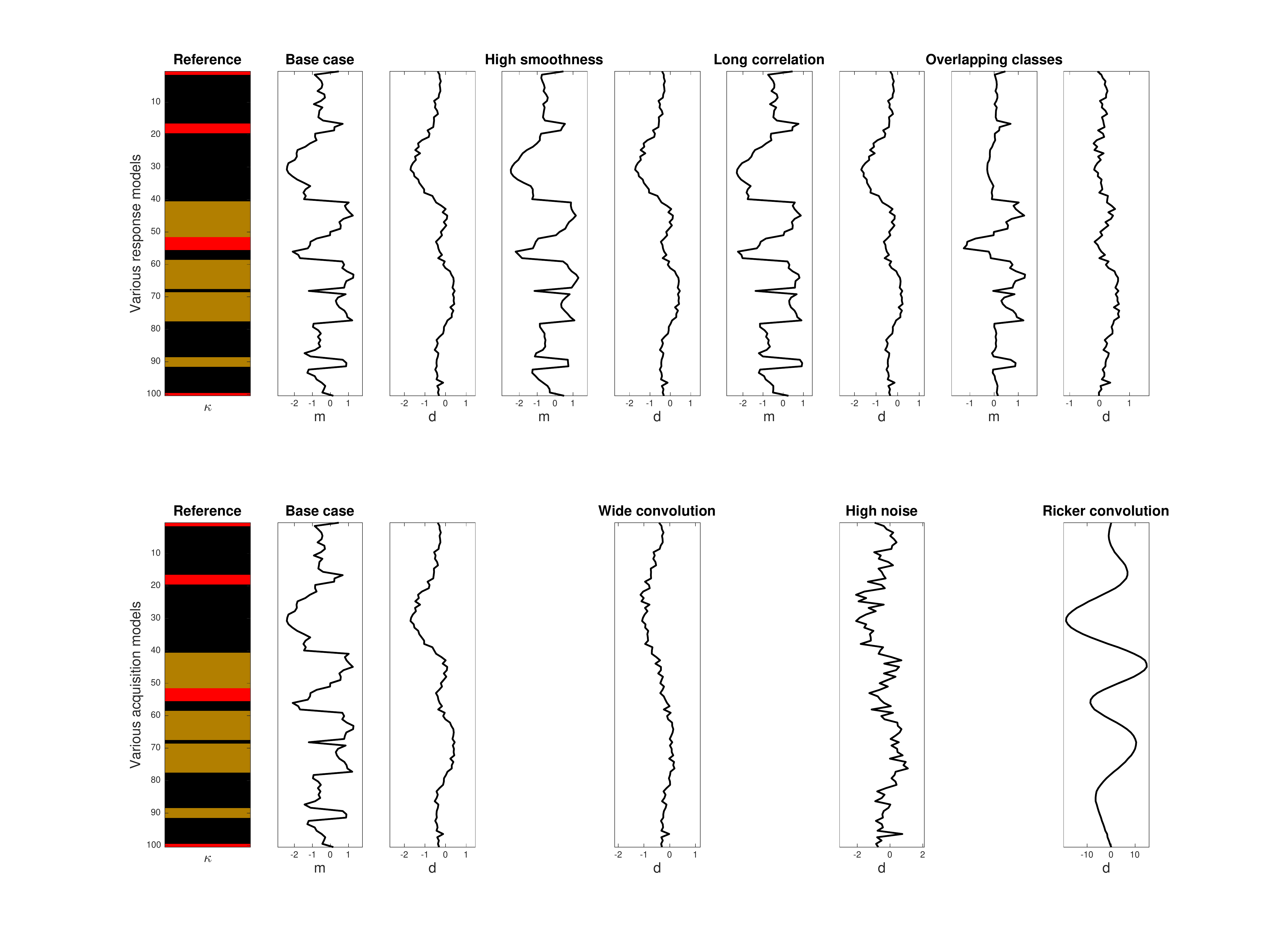}
	\caption{Reference classification $\vkappa^r$, response signals \vm\ and observations \vd\ for the deviating cases. }
		\label{fig:SCS-observations}
\end{figure*}

The reference profile $\vkappa^r$ of interest is assumed to be of length $n = 100$ and have three distinct classes; namely $\{\mathrm{black, \ red, \ brown} \}$, see Figure~\ref{fig:SCS-observations}. The reference profile $\vkappa^r$ is identical for all cases to be discussed, and it is generated as a realization from a first order stationary Markov chain with transition matrix
\begin{equation}
	\Pmat = \begin{pmatrix}
		0.80 & 0.15 & 0.05 \\
		0.15 & 0.75 & 0.10 \\
		0.05 & 0.05 & 0.90
	\end{pmatrix},
\end{equation}
having stationary distribution $p_s = \left( 0.3077, 0.2692, 0.4231\right)\transpose$. 

The Base case has response likelihood model parameters:
\begin{equation}
	\left(\mu_{m_t|\kappa_t},\sigma_{m_t|\kappa_t} \right) = \begin{cases}
		\left(-1, 0.5\right) & \mathrm{if\ black} \\
		\left(0,0.5\right) & \mathrm{if\ red}\\
		\left(1, 0.5\right) & \mathrm{if\ brown}
	\end{cases},
\end{equation}and spatial correlation function $\rhom(h) = \exp \left\{-(h/5)^{1.2} \right\}$ for $h = 0, \ldots, n-1$. The acquisition likelihood model is specified through a second-order exponential convolution kernel, $w(\tau) = (1/\sqrt{12\pi})\times \exp \{-1/2 \times (\tau / 6)^2\}$.

The deviating cases, displayed in Fig.~\ref{fig:SCS-observations}, are defined as follows, relative to the Base case:
\begin{enumerate}
	\item High smoothness: a spatial correlation function with the same range, but a higher degree of smoothness: $\rhom(h) = \exp \left\{-(h/5)^{1.8} \right\}$.
	\item Long correlation: a spatial correlation function with an identical degree of smoothness but a higher range parameter: $\rhom(h) = \exp \left\{-(h/10)^{1.2} \right\}$.
	\item Overlapping classes: the black and red classes have identical pointwise means, but different marginal variances:
\begin{equation}
	\left(\mu_{m_t|\kappa_t},\sigma_{m_t|\kappa_t} \right) = \begin{cases}
		\left(0, 0.1\right) & \mathrm{if\ black} \\
		\left(0,0.5\right) & \mathrm{if\ red}\\
		\left(1, 0.5\right) & \mathrm{if\ brown}
	\end{cases}
\end{equation}
	\item Wide convolution: a second-order exponential convolution kernel: \newline$w(\tau) = (1/\sqrt{24\pi})\times \exp \{-1/2 \times (\tau / 12)^2\}$
	\item High noise: an acquisition likelihood with a large error component: $\Sdgm = 0.5^2\times \Imat{n}$.
	\item Ricker convolution: a convolution kernel with negative lobes: \newline $w(t) \!= \!\left( 1 - 2\pi^2 \times 0.03^2\times t^2\right) \times \exp\{ -\pi^2 \times 0.03^2\times t^2\}$.
\end{enumerate}

Note that cases one through three are defined by different response models, resulting in different \vm\ signals, however they do have an identical acquisition model. Cases four through six have an identical response profile, but different acquisition models. The various set of observations are generated by simulation from the various gross likelihood models given the reference profile $\vkappa^r$.

Fig.~\ref{fig:SCS-observations} contains the reference classification $\vkappa^r$, response signals \vm\ and observations \vd\ displayed for the various cases. The objective is to predict \vkappa\ given \vd\ for each case. The reference profile $\vkappa^r$, and \vm\ and \vd\ in pairs are displayed in the top row for the Base, High smoothness, Long correlation and Overlapping classes cases. Relative to the Base case, \vm\ appears to be smoother in the High smoothness and Long correlation cases, however, the observations appears to be almost identical. For the Overlapping classes case we observe that \vd\ has less variability than in the Base case.

The bottom row displays, in a similar format, the Base case, Wide convolution, High noise and Ricker convolution cases. We do not display \vm\ in the three latter cases since the response models are assumed to be identical to the Base case to ease interpretation. The respective observations are very different, however. 

We assess by simulation the associated signal-to-noise-ratios:
\begin{equation}
	\label{eq:snr}
	\mathrm{snr} = \frac{\mathrm{tr}\left(\Wmat\mumgk \right)}{\mathrm{tr}\left(\Wmat\Srgk\Wmat\transpose + \Sdgm\right)}.
\end{equation}
The signal-to-noise-ratio for the Base case is 2.53, and the signal-to-noise-ratios for the deviating cases are presented in Table~\ref{tab:SCS-snr}. The Base case, High smoothness, Wide convolution and Ricker convolution cases have almost identical signal-to-noise-ratios, while the other cases appear with lower singal-to-noise-ratios. Particularly the High noise case has a large noise component, as it should.

\begin{table}[h!]
	\centering
	\caption{Signal-to-noise-ratios for the various cases.}
		\label{tab:SCS-snr}
		\begin{tabular}{|c|c|c|c|}
		\hline
			\multicolumn{2}{|c|} {Different response models} & \multicolumn{2}{|c|}{Different acquisition models} \\
			\hline
			High smoothness  & 2.49 & Wide convolution & 2.42   \\
			\hline
			Long correlation & 1.71 & High noise & 0.74 \\
			\hline
			Overlapping classes & 1.35 & Ricker convolution & 2.55  \\
			\hline
		\end{tabular}
\end{table}

\begin{figure*}[ht!]
	\centering
	\includegraphics[width =\textwidth]{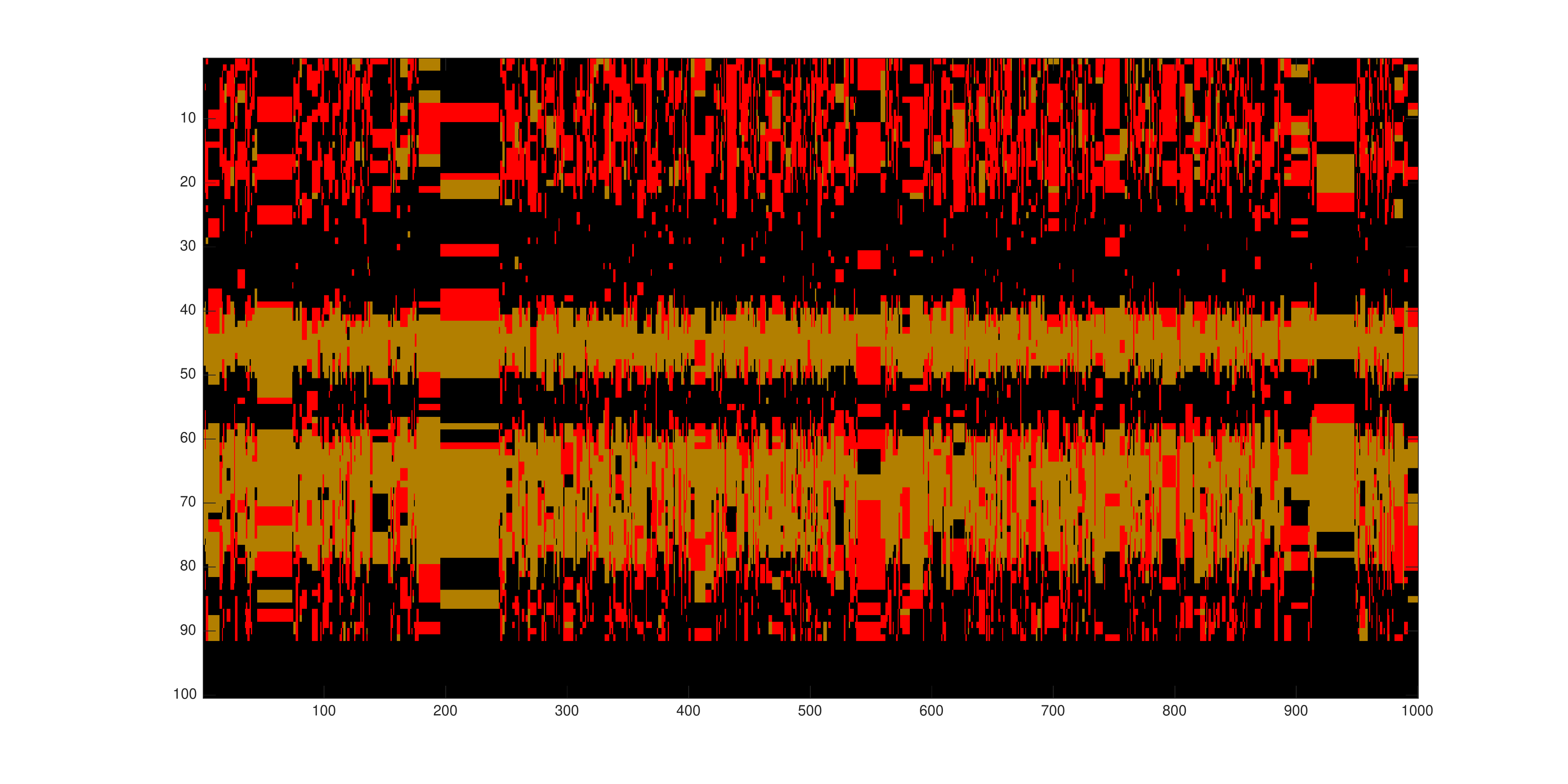}
	\caption{Base case: 1,000 consecutive realizations from the posterior model \cpdf{\vkappa}{\vd} based on a ninth order projection approximation.}
	\label{fig:SCS-samples}
\end{figure*}

For each case we apply both the truncation and the projection approximations to assess the corresponding approximate posterior models by the efficient recursive Forward-Backward algorithm. We use the approximate posterior models as proposal distribution in an independent proposal MCMC MH algorithm to assess the correct posterior model for each case, and 160,000 realizations are generated from the correct posterior model. We discard the initial 10,000 realizations as the burn-in phase.  A higher order delayed rejection step \citep{Trias2009} is included in the algorithm. 
Fig.~\ref{fig:SCS-samples} contains a sequence of 1,000 consecutive posterior realizations for the Base case based on a ninth order projection approximation. We observe relatively good mixing, with the three classes represented in the realizations. This characteristic is shared also by the other cases, but the results are not presented here. 

\begin{figure*}[ht!]
\centering
\includegraphics[width = \textwidth]{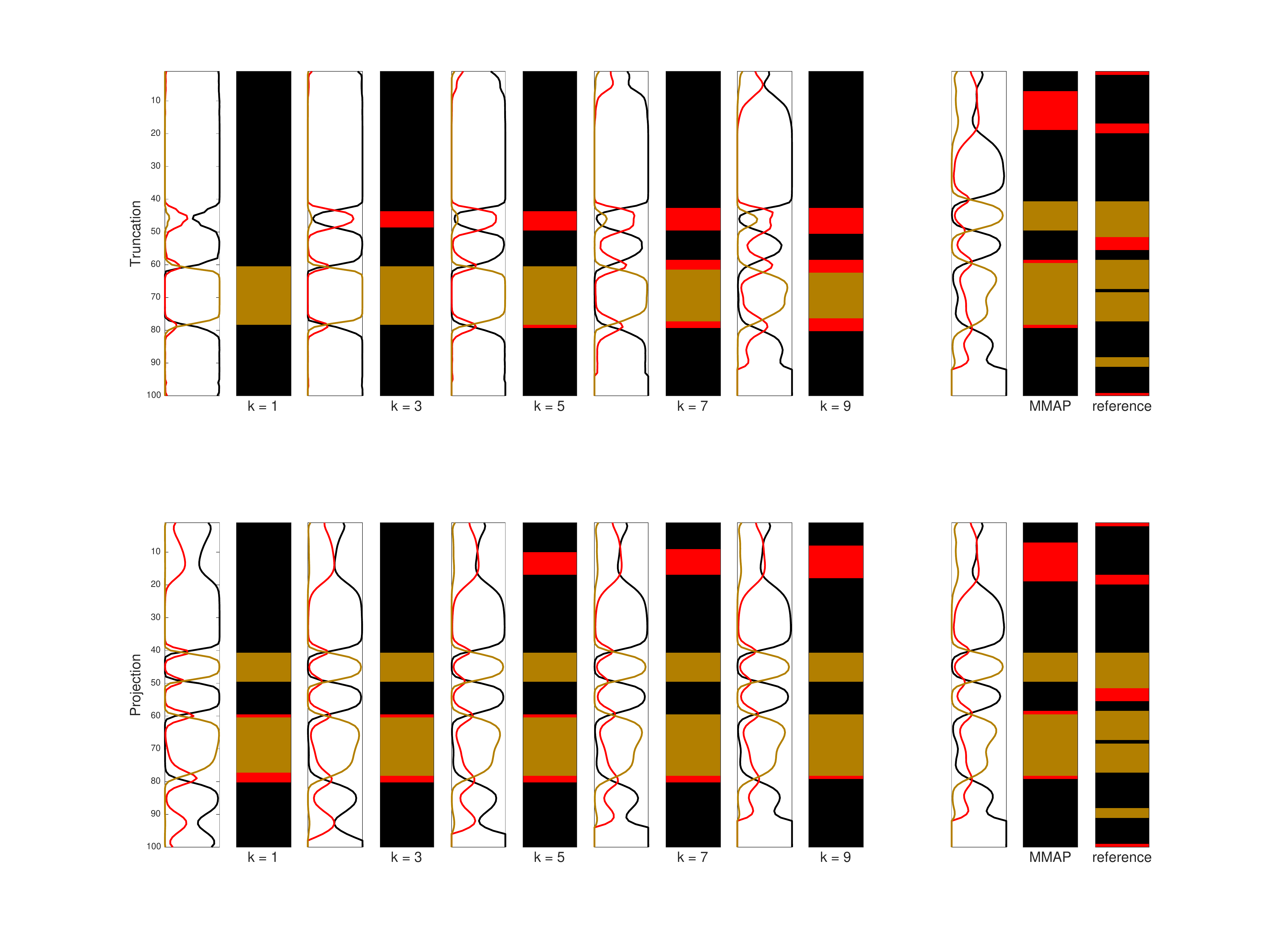}
	\caption{Posterior results Base case. Top row: from left to right are approximate marginal probabilities aMPR$_k^\kappa$ profiles and aMAP$_k$ predictors for the Base case, as functions of $k$ for the truncation approximation. The MPR$_{\kappa}$ profiles are dislayed together with MMAP predictor and reference classification $\vkappa^r$ to the far right. Bottom row: results for the projection approximation in an identical format as the top row.}
	\label{fig:SCS-base-case}
\end{figure*}

In Fig.~\ref{fig:SCS-base-case} results for the Base case are presented. The top line of display is based on the truncation approximation, while the bottom line is for the projection approximation. Both the aMPR$_k^\kappa$ profiles for $\kappa \in \Omega_{\kappa}$ and the aMAP$_k$ predictors are displayed for varying orders of $k$ of approximations. These characteristics are exactly calculated by recursive algorithms, although the computer demand increase fast with increasing order $k$. To the right  estimated MPR$^\kappa$ profiles for $\kappa \in \Omega_\kappa$ and the MMAP predictor for the correct posterior model are displayed together with the reference profile $\vkappa^r$. These characteristics are only available by MCMC based inference. Note that the reference profile $\vkappa^r$ is more heterogeneous than the MMAP predictor due to the regression towards the local majority class.

The aMPR$_k^\kappa$ profiles for the truncation approximation tend towards the MPR$^\kappa$ profiles as the order $k$ increase, which is desirable. It can be shown that for $k=n$ they are identical. It is problematic, however, that for low order $k$ the approximations are not satisfactory, since the computer demands increases fast with $k$. For the projection approximation the results appears to be better; good approximations for low order $k$ and improvements towards the correct model results as $k$ increase. 

Lastly, note that the MCMC estimates of the correct model characteristics are almost identical - independent of whether the proposal is based on the truncation or projection approximation - of course. The MCMC acceptance rates will of course depend on the proposal distribution. 

\begin{figure*}[h!]
	\centering
\includegraphics[width = 1\textwidth]{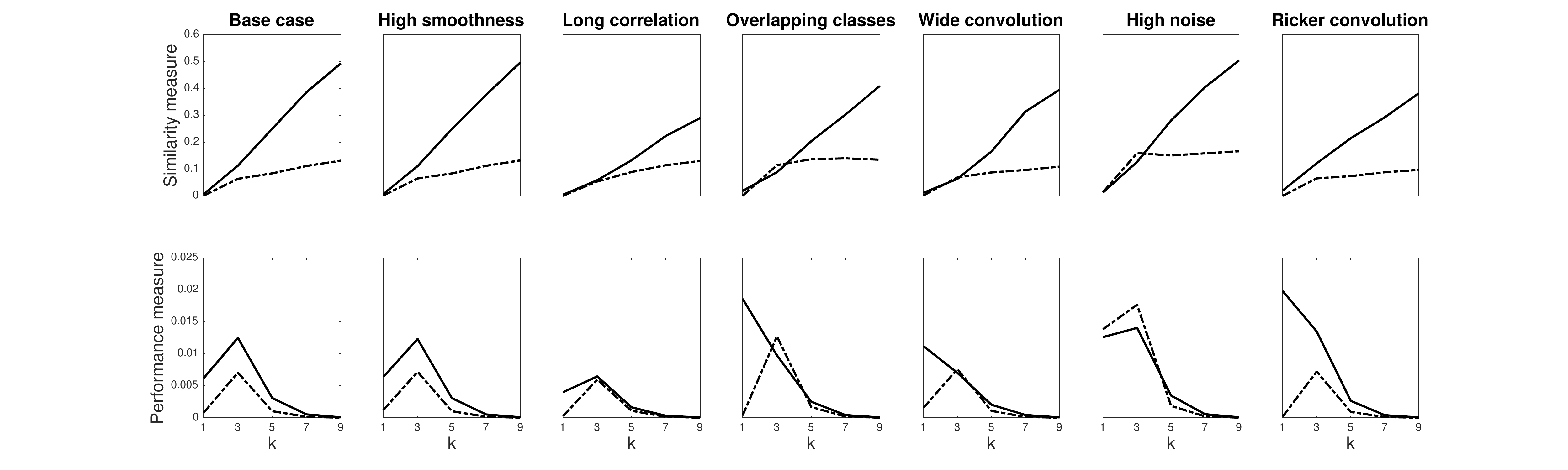}
	\caption{Top row: approximate similarity measures $\alpha_{t}^{(\kk)}$ and $\alpha_{p}^{(\kk)}$ based on the truncation (dashed line) and projection (solid line) approximations. Bottom row: performance measures $\beta^{(k)}$ based on the truncation (dashed line) and projection (solid line) approximations.}
		\label{fig:SCS-acceptance-rates}
\end{figure*}

In Fig.~\ref{fig:SCS-acceptance-rates} the similarity measures $\alpha_t^{(k)}$ and $\alpha_p^{(k)}$ for the approximate versus correct posterior model for all cases are presented in the top row for increasing $k$. The corresponding performance measures $\beta_t^{(k)}$ and $\beta_p^{(k)}$, taking also computer demands into account, are displayed in the bottom row.

The similarity measures $\alpha_t^{(k)}$ and $\alpha_p^{(k)}$ appears to increase monotonically with increased order $k$ in the various cases. This indicates that the two sequences of approximations, parametrized by order $k$, provides monotonical improved approximations for the correct posterior model with increasing $k$ at the cost of increased computer demands. The projection approximation is almost uniformly better than the truncation approximation. For all cases, the similarity measure for the former reaches the range $0.3 - 0.5$ for a ninth order approximation, which entails average acceptance rates of $30-50$ \% in the independent proposal MCMC MH algorithm. These acceptance rates are very satisfactory, but the computer demands for the ninth order projection approximation is large. In the bottom line, we observe that the performance measure indicates that the order $k$ in around 3 is optimal with acceptance rates in range $10-20$ \%.

\begin{figure*}[ht!]
\includegraphics[width = \textwidth]{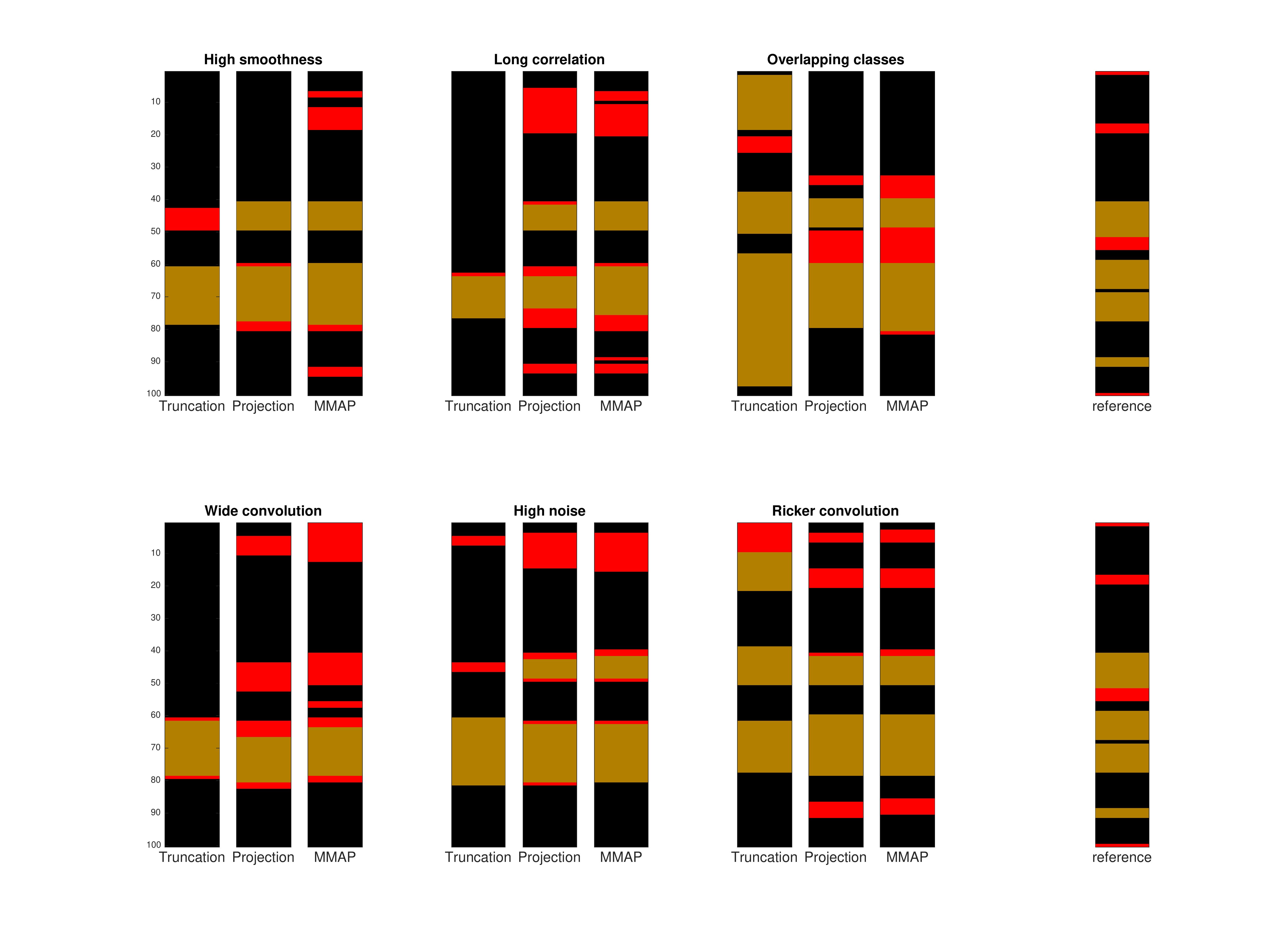}
	\caption{Posterior results deviating cases. Each triplet includes, from left to right, aMAP$_9$ predictors based on the truncation and projection approximations, and MMAP predictor for the correct posterior model for all cases. To the far right in each row is the reference profile $\vkappa^r$ displayed.}
		\label{fig:SCS-MAP}
\end{figure*}

Fig.~\ref{fig:SCS-MAP} contains exact aMPR$_k^\kappa$ profiles for the deviating cases for $\kappa \in \Omega_\kappa$ and exact aMAP$_k$ predictors for the truncation and projection approximations of order $k = 9$. Also, the estimated MPR$^\kappa$ profiles for $\kappa \in \Omega_\kappa$ and MMAP predictor for the correct posterior models are displayed. The aMPR$_{k}^\kappa$ profiles and aMAP$_k$ predictors based on the projection approximation are closer to the MRP$^\kappa$ profiles and MMAP predictors for the correct posterior model than the ones based on the truncation approximation. It is particularly so for the Long correlation, Overlapping classes, Wide convolution and Rick convolution cases, that is, for cases with strong spatial coupling between states. The aMPR$_k^\kappa$ profiles for $\kappa \in \Omega_\kappa$ and aMAP$_k$ predictors, which can be exactly assessed by recursive algorithms, are so close to the corresponding characteristics for the correct posterior model that one may question the necessity of a MCMC step. Recall that calculating the posterior density $\posterior$ in each MCMC iteration is computer demanding. 

\FloatBarrier

\begin{figure}[h!]
	\centering
	\includegraphics[height = \textheight]{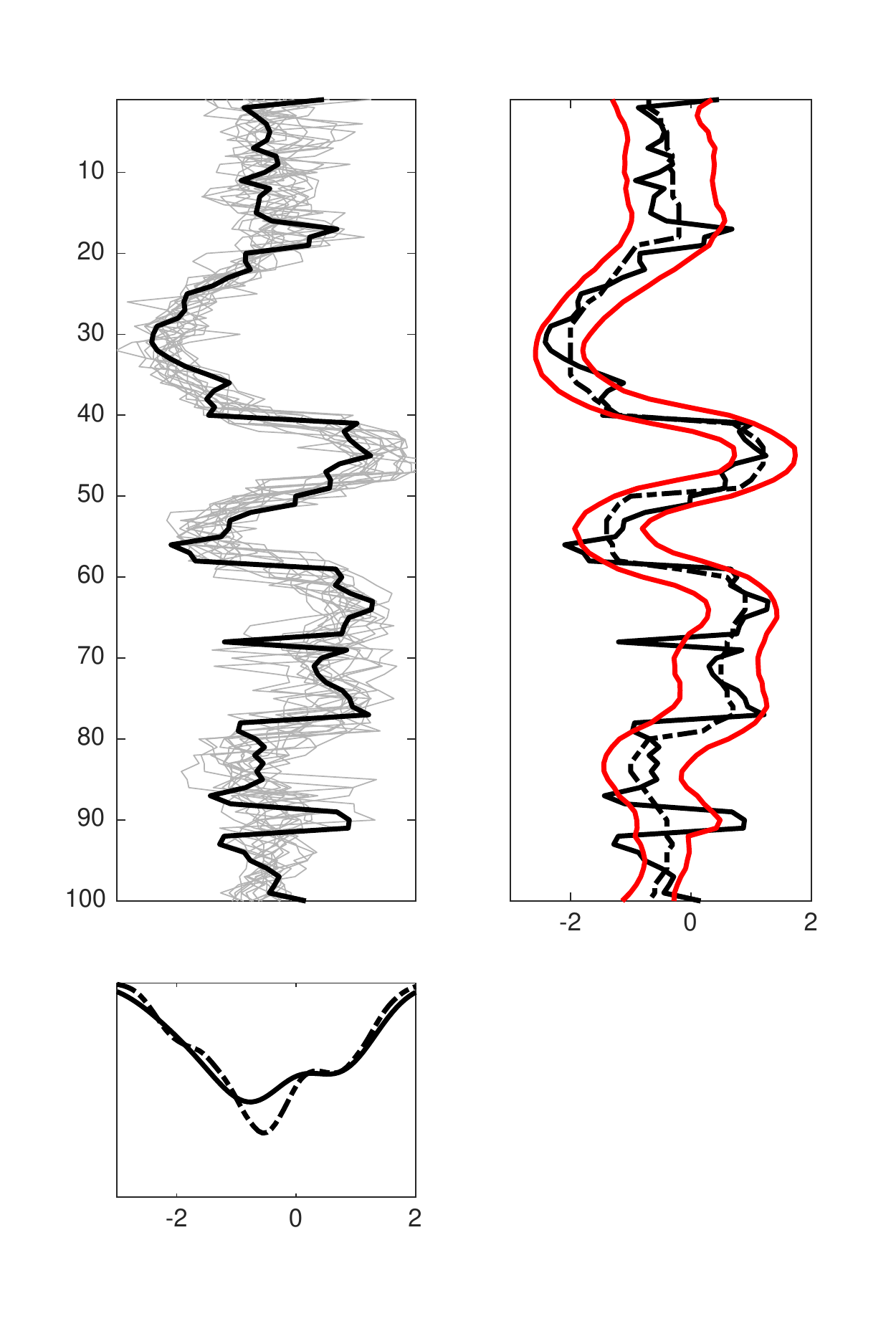}
	\caption{Posterior results Base case: Top row: true response profile $\vm$ (solid black) together with conditional realizations from $\cpdf{\vm}{\vd}$, and true response profile $\vm$ (solid black line), MMAP predictor $\hat{\vm}$ (dashed black line) and 80 \% prediction interval (solid red lines). Bottom row: fitted densities $\pdf{\vm}$ (solid line) and $\cpdf{\vm}{\vd}$ (dashed line).}
	\label{fig:BC-m}
\end{figure}

\begin{figure}[h!]
	\centering
	\includegraphics[width =.7\textwidth]{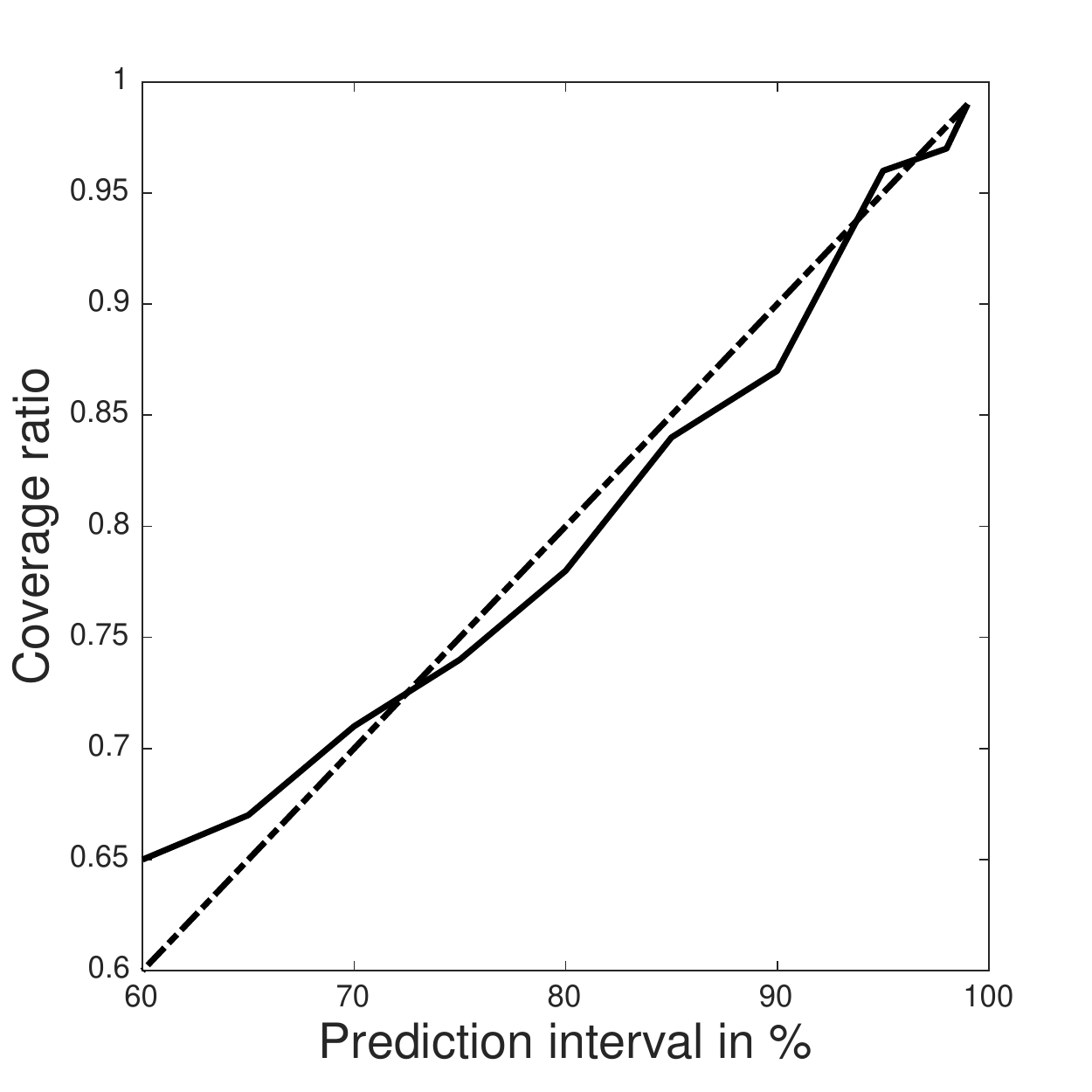}
	\caption{Base case empirical coverage ratio for the prediction interval (solid line) as a function of width of the prediction interval (in \%) together with reference coverage ratio (dashed line).}
	\label{fig:coverage_ratio}
\end{figure}

We define the following MMAP predictor: $\hat{\vm} = \left( \hat{m}_1,\ldots, \hat{m}_n\right)\transpose$ from Eq.~\eqref{eq:posterior-mt}. Posterior realizations, MMAP predictor, 80 \% prediction interval and the fitted density \cpdf{\vm}{\vd} are displayed in Fig.~\ref{fig:BC-m}. We observe that the MMAP predictor $\hat{\vm}$ captures the class transitions in $\vkappa$. The fitted density $\cpdf{\vm}{\vd}$ appear with similar characteristics, such as bimodality and skewness, as \pdf{\vm}. Compared to the true response profile \vm\ we obtain a root mean squared error (rmse) value of 0.54 in the Base case. In Fig.~\ref{fig:coverage_ratio} we display the empirical coverage ratios for various prediction interval, and we observe the coverage ratios to be satisfactory.

\begin{figure}[h!]
	\centering
	\includegraphics[width =\textwidth]{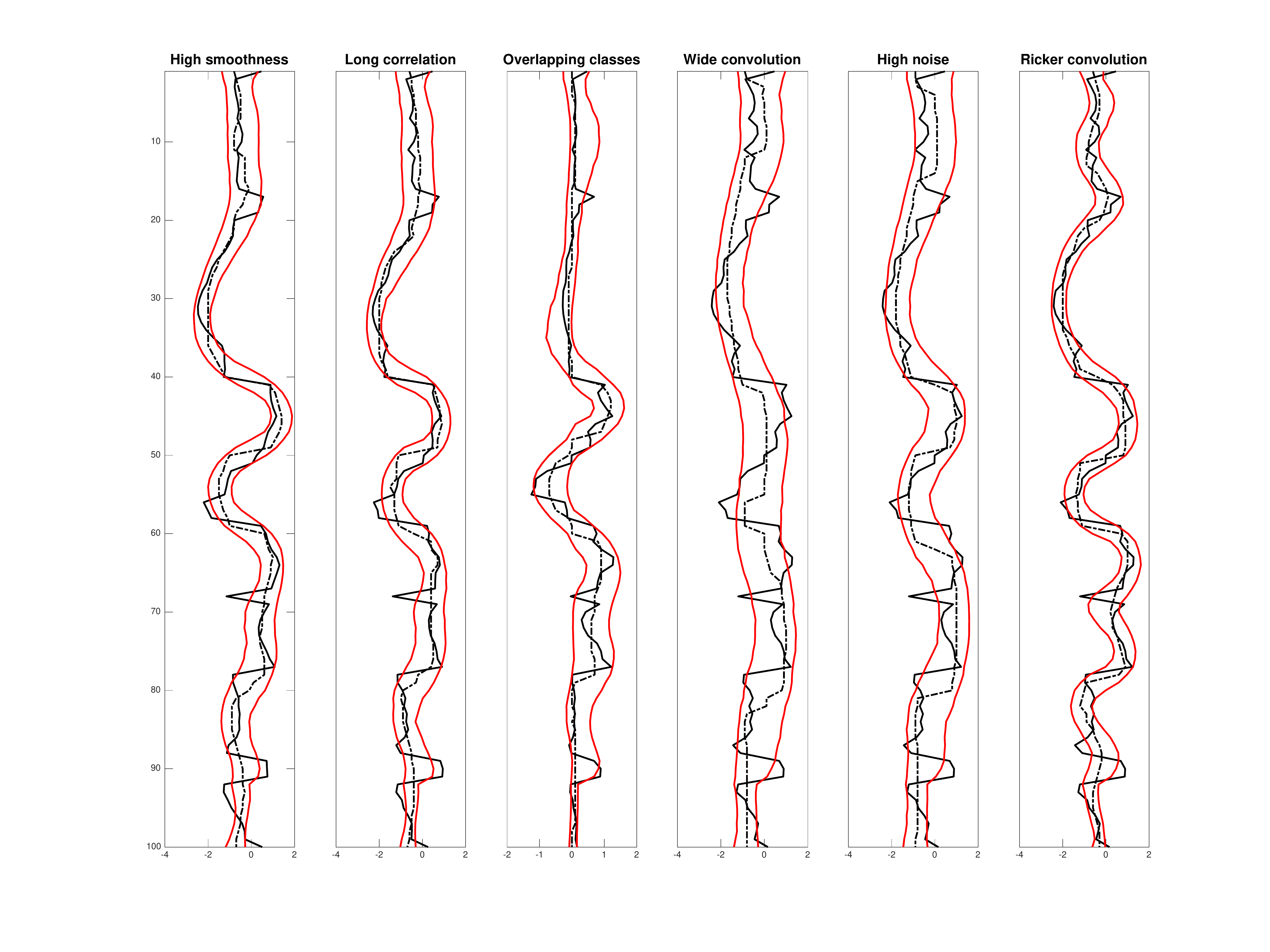}
	\caption{Posterior results deviating cases. True response profile $\vm$ (solid black line), MMAP predictor $\hat{\vm}$ (dashed black line) and 80 \% prediction interval (solid red lines) for the deviating cases.}
	\label{fig:BC-m-all}
\end{figure}

In Fig.~\ref{fig:BC-m-all} we present posterior results for $\cpdf{\vm}{\vd}$ for the deviating cases. The MMAP predictors are seen to capture the main characteristics of the true response profile \vm\ in most of the cases. Also, the 80 \% prediction intervals appear to be satisfactory. Coverage ratios for the 80 \% prediction intervals and  rmse values are given in Tab.~\ref{tab:rmse} for the deviating cases. As expected, the rmse values are higher in the Wide convolution and High noise case. Somewhat surprisingly the rmse values are almost identical in the Base, High smoothness, Long correlation and Ricker convolution cases. In general, the 80 \% coverage ratios are satisfactory.  

\begin{table}[h!]
	\centering
	\caption{Root mean squared errors and coverage ratios for 80 \% prediction intervals.}
		\label{tab:rmse}
		\begin{tabular}{|c|c|c|}
		\hline
			& Root mean squared error & Coverage ratio \\
			\hline
			High smoothness  & 0.56 & 77 \% \\
			\hline
			Long correlation & 0.52 & 79 \% \\
			\hline
			Overlapping classes &0.29 & 89 \% \\
			\hline
			Wide convolution & 0.83& 70 \% \\
			\hline
			High noise & 0.75  & 76 \% \\
			\hline
			Ricker convolution & 0.53 & 78 \% \\
			\hline
		\end{tabular}
\end{table}


To conclude - the projection approximation is clearly preferable to the truncation approximation, particularly for models where the likelihood functions have a high degree of spatial coupling. These models are indeed the ones most challenging to invert. For hard problems with many classes a projection approximation of order $k = 3$ and a MCMC step to assess the correct posterior model is recommended. In most cases with a small to moderate number of classes a projection approximation of order $k$ about nine will provide aMPR$_k^\kappa$ profiles for $\kappa \in \Omega_\kappa$ and aMAP$_k$ predictors that are so close to the correct MPR$^\kappa$ profiles and MMAP predictor, that the MCMC step need not be necessary. The MMAP predictor $\hat{\vm}$ is seen to capture the main characteristics of the true response profile $\vm$, and the coverage ratios appear to be satisfactory compared to the reference. Bimodality and skewness are present in the posterior model \cpdf{\vm}{\vd}, as desired.

\section{Norwegian Sea case study}
\label{sec:RCS}
Lithology/fluid prediction subsurface is a major challenge in reservoir characterization. The variables of interest are the lithology/fluid classes and elastic attributes along a vertical profile penetrating a reservoir unit, and the objective is to predict these variables given a set of seismic amplitude-versus-observations (AVO) observations. We demonstrate the proposed methodology on a case study from the Norwegian Sea. The data set covers a mid-Jurassic gas sandstone reservoir in the Garn and Ile formation, separated by a thick silty-shale formation. The reservoir zones are characterized by a relatively low P-wave velocity and density. We refer to \cite{Avseth2016} for further details about the reservoir of interest.

\begin{figure}[h!]
	\centering
	\includegraphics[width = \textwidth]{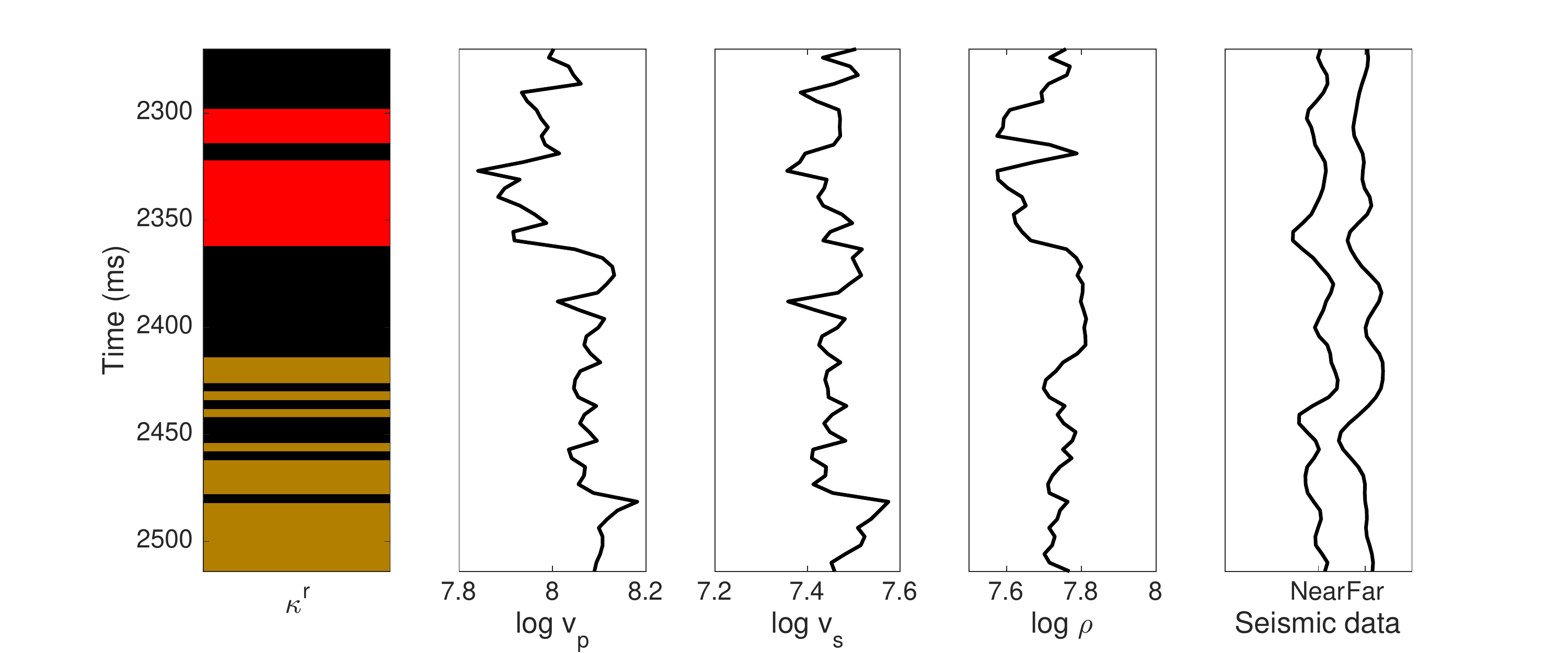}
	\caption{Geologically interpreted reference lithology/fluid classes $\vkappa^r$ (shale in black, gas sandstone in red and brine sandstone in brown),  elastic material attributes $\left(\log v_p, \log v_s, \log \rho \right)$ and observed seismic AVO data along the well profile.}
	\label{fig:CS-ref}
\end{figure}

We discretize the upscaled region of interest onto a regular grid of size $n = 61$, and we operate in time-domain. Three distinct lithology/fluid classes are identified in a reference solution $\vkappa^r$ (Fig.~\ref{fig:CS-ref}), namely, $\Omega_{\kappa} = \{\mbox{shale, gas sandstone, brine sandstone}\}$. We define a first order Markov chain prior model for the lithology/fluid classes: 
\begin{equation}
	\Pmat = \begin{pmatrix}
		0.75 & 0.05 & 0.20 \\
		0.15 & 0.85 & 0 \\
		0.30 & 0 & 0.70 
	\end{pmatrix},
\end{equation}
having stationary distribution $\left(0.50, 0.17, 0.33\right)\transpose$. We note the prior zero-probability transitions between gas sandstone and brine sandstone due to gravitational sorting \citep{Krumbein1969}.

\begin{figure}[h!]
	\centering
	\includegraphics[width = \textwidth]{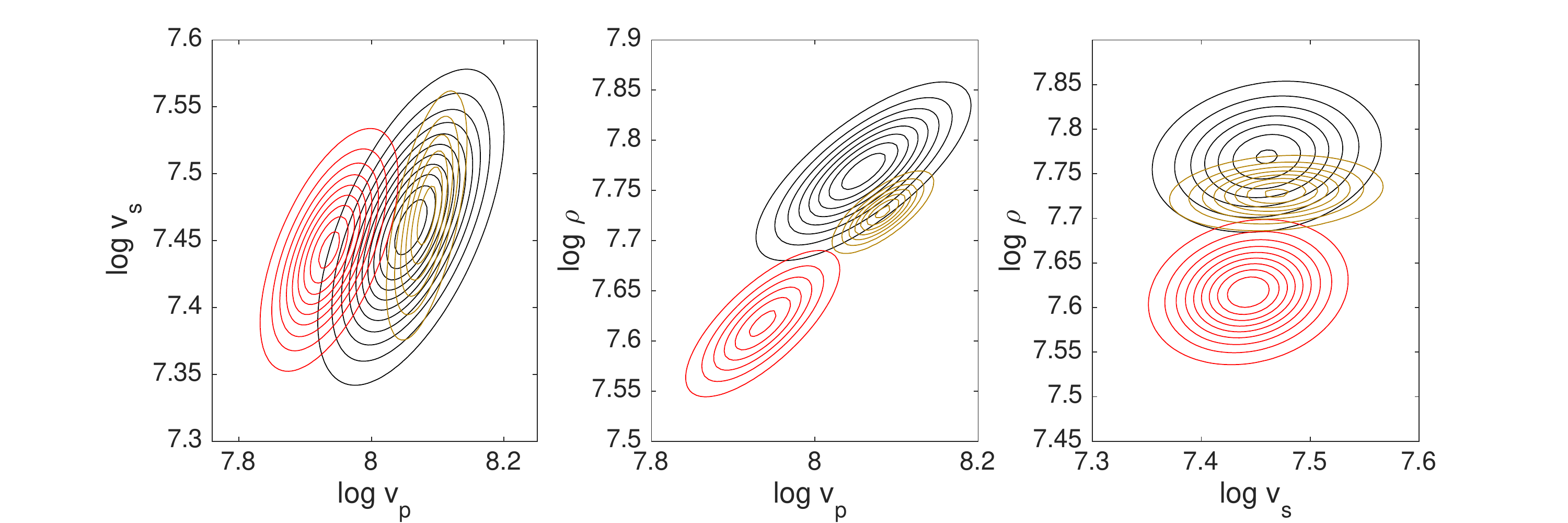}
	\caption{Contour plots of empirically calibrated rock-physics likelihood models $\cpdf{m_t}{\kappa_t}$. Colour coded by lithology/fluid class: shale (black), gas sandstone (red) and brine sandstone (brown).}
	\label{fig:CS-rpm}
\end{figure}

The middle-layer \vm\ represents the logarithm of the elastic material properties,  which are the canonical variables for the seismic AVO observations. These properties are parametrized by the logarithm of pressure-wave velocity $\log v_p$, the logarithm of shear-wave velocity $\log v_s$, and the logarithm of the density $\log \rho$. The response likelihood model, or rock-physics model \citep{Avseth2005, Mavko2009}, is empirically calibrated from an upscaled nearby well. Contour plots for the various $\cpdf{m_t}{\kappa_t}$ are displayed in Fig.~\ref{fig:CS-rpm}. Indeed, as seen in the contour plot, the elastic attributes are correlated. An exponential spatial correlation function $\rhom(h)$ is also estimated from the nearby well.

The upper-layer \vd\ represents angle-dependent seismic AVO data from near and far angles (Fig.~\ref{fig:CS-ref}). The reflections and wave propagation subsurface are modeled by a convolutional, linearized Zoeppritz version of the wave equation \citep{Buland2003}. That is, $\Wmat  = \mat{CAD}$, where $\mat{C}$ is a convolutional matrix, \Amat\ is the angle-dependent weak-contrast Aki-Richards coefficients \citep{Aki1980}, and $\Dmat$ is a differential matrix which calculates contrasts. The covariance matrix in the acquisition model is assumed to be $\Sdgm = 300 \times \Imat{2n}$. The signal-to-noise-ratio is assessed using Eq.~\eqref{eq:snr}, and it is estimated to be 2.23, which is comparable to most cases discussed in Section~\ref{sec:RCS}.

We consider only the projection approximation for $k \in \{2,\ldots, 7\}$ based on the discussion in Section~\ref{sec:Synthetic}. A set of 100,000 realizations is generated from \posterior, and the initial 10,000 realizations are discarded as the burn-in phase. The similarity measures $\alpha_p^{(k)}$ and the performance measures $\beta_p^{(k)}$ are given in Table~\ref{tab:CS-acceptance}. We obtain acceptance rates in the range 0.15 - 0.35, which is monotonically increasing with $k$ except for $k = 6$. As in Section~\ref{sec:RCS}, $k$ around 2-3 appear to be a reasonable trade-off between acceptance rate and computational cost.

\begin{table}
	\centering
		\caption{Similarity measure $\alpha_p^{(k)}$ and performance measure $\beta_p^{(k)}$ for case study.}
	\label{tab:CS-acceptance}
	\begin{tabular}{|p{2.4cm}|p{1cm}|p{1cm}|p{1cm}|p{1cm}|p{1cm}|p{1cm}|p{1cm}|}
		\hline
		& $\!k= 2\!$ & $\!k= 3\!$ & $\!k= 4\!$ & $\!k= 5\!$ & $\!k= 6\!$ &$\!k= 7\!$ \\ 
		\hline
		Similarity \newline measure & 0.1612 & 0.2139 & 0.2580& 0.2812& 0.2392&0.3460\\
		\hline
		Performance \newline measure &   0.0537   & 0.0238 &   0.0096 &  0.0035  &  0.0010  & 0.0005\\
	\hline
	\end{tabular}
\end{table}

\begin{figure}[h!]
	\centering
	\includegraphics[width = \textwidth]{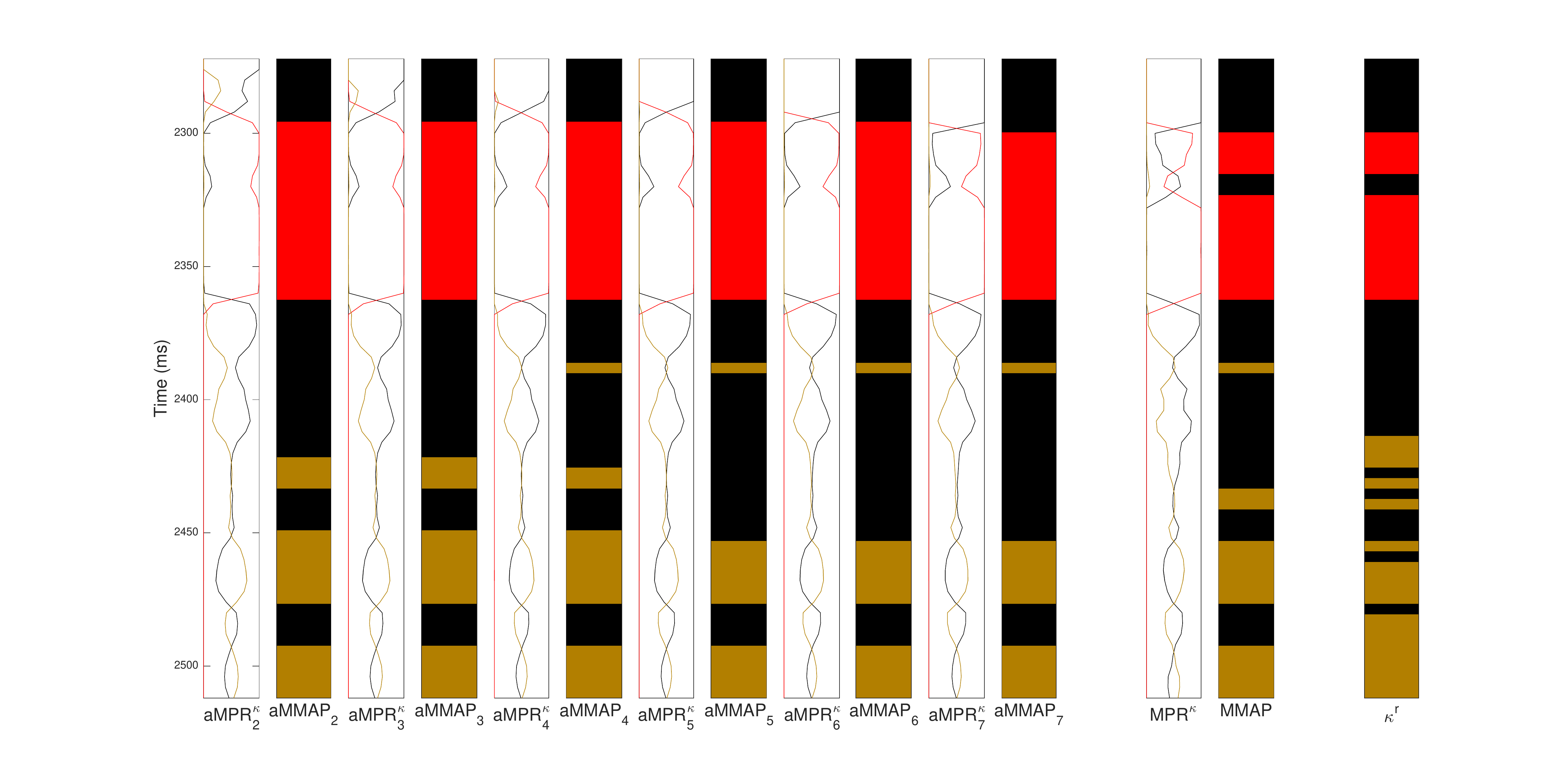}
	\caption{From left to right are, in pairs, aMPR$_k^\kappa$ profiles and aMMAP$_k$ predictor for $k = 2,\ldots, 7$, MPR$^\kappa$ profiles and MMAP predictor, and reference profile $\vkappa^r$.}
	\label{fig:CS-posterior-kappa}
\end{figure}

Fig.~\ref{fig:CS-posterior-kappa} contains aMPR$_k^\kappa$ profiles for $\kappa \in \Omega_{\kappa}$ and aMMAP$_k$ predictors for $k = 2,\ldots, 7$ based on the projection approximation. Both the aMPR$_k^\kappa$ profiles and the aMMAP$_k$ predictors appear to be similar and almost identical for $k \geq 5$. The estimated MPR$^{\kappa}$ profiles are displayed together with the MMAP predictor, and we observe that most of their characteristics are shared by the approximate solutions. However, we note that in contrast to the aMMAP$_k$ predictors, the thin silty-shale layer around 2320 ms is identified in the MMAP. The brine sandstone layer at 2430 ms is captured in the aMAP$_k$ for low order $k$, but it is not captured for higher order $k$. As seen in the aMPR$_k^\kappa$ profiles the marginal probability for brine sandstone is approximately 50 \% around 2430 ms, thus it does not influence the proposal density in the MCMC algorithm significantly. The MMAP predictor is observed to capture the main characteristics of the reference profile $\vkappa^r$, however small-scale variability is lost since predictions causes a regression towards the dominating class; shale. We observe that the MMAP predictor is more uncertain in the lowermost part of the the domain of interest based on the MPR$^{\kappa}$ profiles. Indeed, this in correspondence with the rock-physics model where we see that the brine sandstone is partly masked by shale whereas gas sandstone is clearly separated from shale (Fig.~\ref{fig:CS-rpm}). 

\begin{figure}[h!]
	\centering
	\includegraphics[width = \textwidth]{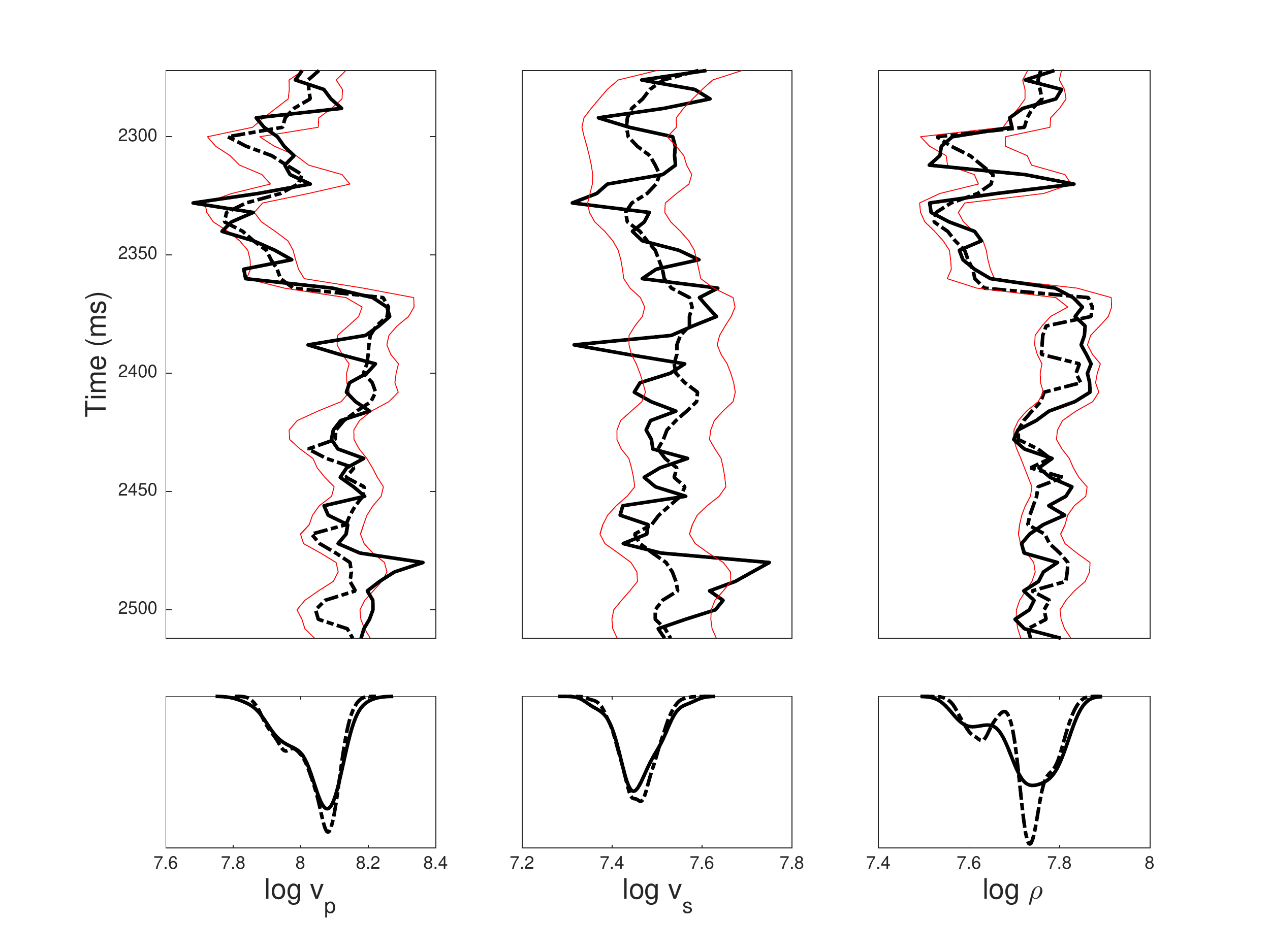}
	\caption{Top row: posterior results for the elastic properties: observed elastic material properties (black solid line), MMAP predictor (black dashed line) and 80 \% prediction interval (red solid line). Bottom row: fitted marginal densities for well observations (solid line) and posterior model (dashed line).}
	\label{fig:CS-posterior-m}
\end{figure}

Posterior predictions and prediction intervals for the elastic attributes are displayed in Fig.~\ref{fig:CS-posterior-m}. In specific, we observe that we are able to predict low $\log v_p$ and $\log \rho$ zones where the reservoirs are located. We display the kernel smoothed time-averaged histograms at the bottom in Fig.~\ref{fig:CS-posterior-m} and observe the posterior models for $\log v_p$ and $\log v_s$ to resemble the observed smoothed histograms based on the well observations. Root mean squaredd errors for the predictions are $\left(0.0414, 0.0423, 0.0391\right)$ for $\left(\log v_p,\log v_s,\log \rho\right)$. Coverage ratios for the 80 \% prediction intervals are $\left(72.1 \% , 77.0\%, 86.9 \%\right)$ for $\left(\log v_p,\log v_s,\log \rho\right)$, which are satisfactory.

\begin{figure}[h!]
	\centering
	\includegraphics[width =\textwidth]{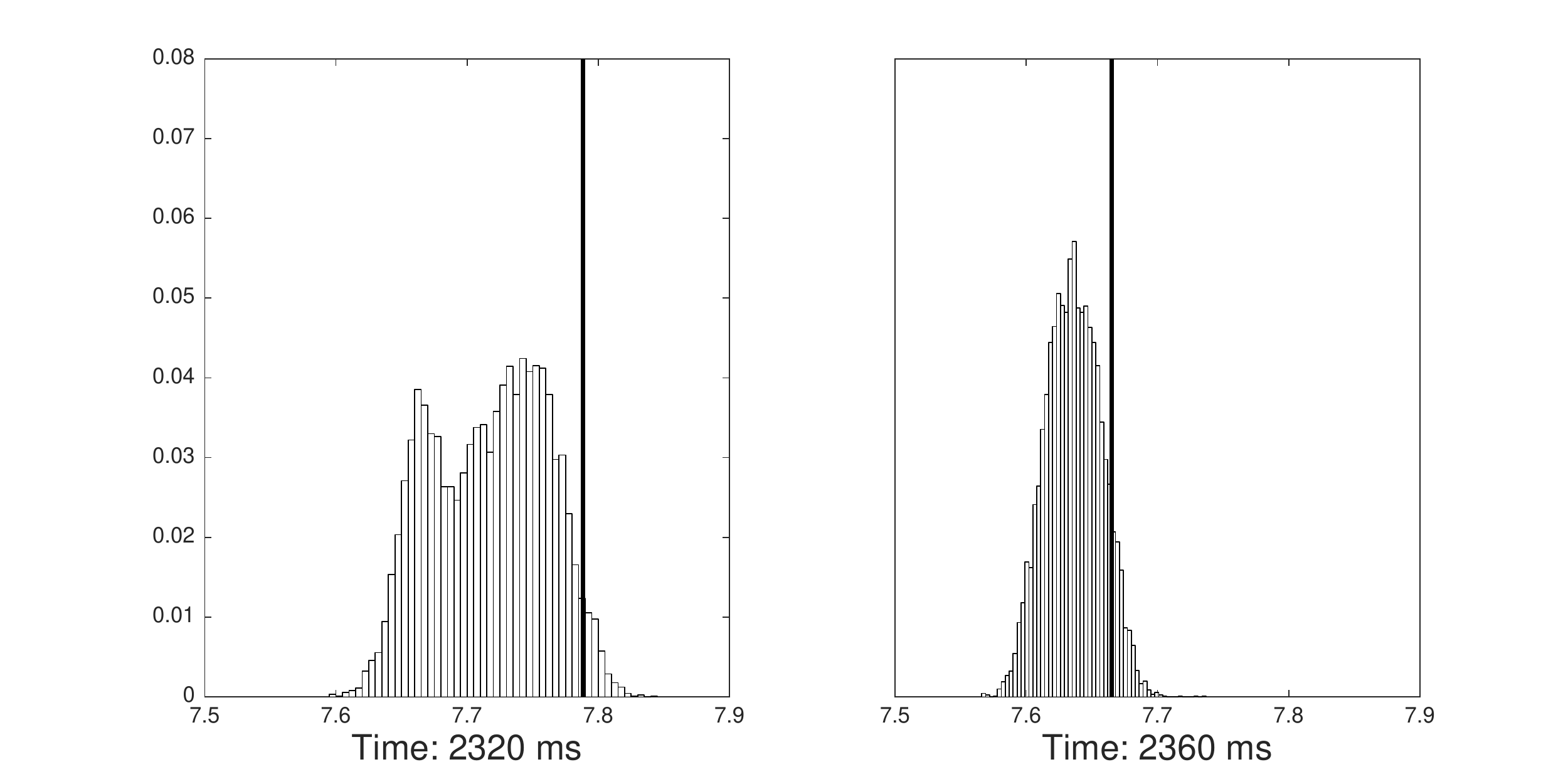}
	\caption{Histogram of $[\log \rho|\vd]$ at 2320 ms and 2360 ms. Correct well observation is displayed with a solid line.}
	\label{fig:CS-hist-m}
\end{figure}

As seen in Fig.~\ref{fig:CS-posterior-m} the width of the prediction intervals varies with time, in specific, the width of the prediction interval in the uppermost gas reservoir is wider than in the lowermost gas reservoir. Fig.~\ref{fig:CS-hist-m} displays histograms of $[\log \rho|\vd]$ at 2320 ms and 2360 ms. The posterior is observed to wider at the former. At 2320 ms the posterior is bimodal, while it is unimodal at 2360 ms. This is in consistent with the MPR$^\kappa$ profiles displayed in Fig.~\ref{fig:CS-posterior-kappa} where the marginal probability for shale is higher in the upper reservoir. That is, the uncertainty in the lithology/fluid classification propagates into the uncertainty of the posterior for the elastic properties.

\FloatBarrier

\section{Conclusion}
A convolved hidden Markov model extending \cite{Larsen2006} and \cite{Rimstad2013} is defined inspired by applications in subsurface reservoir prediction. The bottom-layer is a categorical model, and the middle-layer is Gaussian spatial model conditional on the bottom-layer. Hence, the middle-layer appears as an unconditional Gaussian mixture spatial variable. The top-layer, representing the observations, is assumed to be Gaussian conditional on the middle-layer. Prediction of the categorical and Gaussian mixture variables, given the observations in a Bayesian inversion framework, is of interest. 

Two classes of likelihood approximations are presented to obtain approximate posterior models on factorial form, which are analytically and efficiently assessed by the recursive Forward Backward algorithm. The first approximation is based on a na\"{\i}ve truncation of the marginal densities, while the second approximation is based on a Gaussian approximation of a Gaussian mixture density. The correct posterior model is thereafter assessed using an approximate posterior model as proposal density in an independent proposal MCMC MH algorithm. The approximations are empirically evaluated on a set of synthetic cases. In general, higher order approximations results in acceptance rates up to around 0.5, at the cost of additional computational resources which increases exponentially. The projection approximation appears with higher acceptance rates in the MCMC algorithm than the truncation approximation. We observe that $k$ about two or three appears to be a reasonable trade-off between accuracy and computational cost. 

The MMAP predictor for the middle-layer is verified to reproduce bimodality and skewness in the middle layer in the synthetic examples. Coverage ratios for the prediction intervals are observed to be satisfactory.

The projection approximation have been empirically verified on a subsurface case study to predict both the lithology/fluid classes and elastic material properties reliably. 

We observe that for a higher order approximation additional MCMC sampling may not be necessary since the approximate posterior models appear to be very close to the correct posterior model.

The current work should naturally be extended to 2D and 3D models along the lines in \cite{Ulvmoen2010} and \cite{Rimstad2010}. Preliminary work on real 2D subsurface seismic show encouraging results.

\subsection*{Acknowledgments}
This research is part of the Uncertainty in Reservoir Evaluation (URE) activity at the Norwegian University of Science and Technology (NTNU). We thank PGS and Tullow Oil for providing the data. 


\bibliographystyle{chicago}
\bibliography{references3}

\end{document}